\input harvmac
\def\np#1#2#3{Nucl. Phys. {\bf B#1} (#2) #3}
\def\pl#1#2#3{Phys. Lett. {\bf #1B} (#2) #3}

\def\pr#1#2#3{Phys. Rev. {\bf #1} (#2) #3}
\def\ap#1#2#3{Ann. Phys. {\bf #1} (#2) #3}

\def\cmp#1#2#3{Comm. Math. Phys. {\bf #1} (#2) #3}
\def\mpl#1#2#3{Mod. Phys. Lett. {\bf #1} (#2) #3}

\def\jhep#1#2#3{JHEP {\bf#1}(#2) #3}

\def\ijmp#1#2#3{Int.~J.~Mod.~Phys. {\bf #1} (#2) #3}
\def\atmp#1#2#3{Adv.~Theor.~Math.~Phys.{\bf #1} (#2) #3}
\def\ap#1#2#3{Ann.~Phys. {\bf #1} (#2) #3}
\def\IB{\relax\hbox{$\inbar\kern-.3em{\rm B}$}}
\def\IC{\relax\hbox{$\inbar\kern-.3em{\rm C}$}}
\def\ID{\relax\hbox{$\inbar\kern-.3em{\rm D}$}}
\def\IE{\relax\hbox{$\inbar\kern-.3em{\rm E}$}}
\def\IF{\relax\hbox{$\inbar\kern-.3em{\rm F}$}}
\def\IG{\relax\hbox{$\inbar\kern-.3em{\rm G}$}}
\def\IGa{\relax\hbox{${\rm I}\kern-.18em\Gamma$}}
\def\IH{\relax{\rm I\kern-.18em H}}
\def\IK{\relax{\rm I\kern-.18em K}}
\def\IL{\relax{\rm I\kern-.18em L}}
\def\IP{\relax{\rm I\kern-.18em P}}
\def\IR{\relax{\rm I\kern-.18em R}}
\def\IZ{\relax\ifmmode\mathchoice{
\hbox{\cmss Z\kern-.4em Z}}{\hbox{\cmss Z\kern-.4em Z}}
{\lower.9pt\hbox{\cmsss Z\kern-.4em Z}}
{\lower1.2pt\hbox{\cmsss Z\kern-.4em Z}}
\else{\cmss Z\kern-.4em Z}\fi}
\def\II{\relax{\rm I\kern-.18em I}}

\def\ndt{{\noindent}}

\def\ee#1{{\rm erf}\left(#1\right)}
\def\sssec#1{\ndt$\underline{#1}$}

\def\CA{{\cal A}}

\def\CD{{\cal D}}
\def\CE{{\cal E}}

\def\CG{{\cal G}}
\def\CH{{\cal H}}

\def\CL{{\cal L}}

\def\CN{{\cal N}}

\def\CT{{\cal T}}

\def\CX{{\cal X}}

\def\p{\partial}
\def\pb{\bar{\partial}}


\def\tb{\bar{t}}
\def\zb{\bar{z}}


\def\Tr{{\rm Tr}}
\def\Id{{\rm Id}}


\def\inbar{\,\vrule height1.5ex width.4pt depth0pt}

\font\cmss=cmss10 \font\cmsss=cmss10 at 7pt

\def\a{{\alpha}}

\def\d{{\delta}}

\def\e{{\epsilon}}
\def\z{{\zeta}}
\def\ve{{\varepsilon}}
\def\vf{{\varphi}}
\def\m{{\mu}}
\def\n{{\nu}}
\def\u{{\Upsilon}}
\def\l{{\lambda}}
\def\s{{\sigma}}
\def\t{{\theta}}

\def\lref{\begingroup\obeylines\lr@f}
\def\lr@f#1#2{\gdef#1{\ref#1{#2}}\endgroup\unskip}
\lref\mtoda{P.~Etingof, I.~Gelfand, V.~Retakh,
``Factorization of differential operators, quasideterminants, and
nonabelian Toda field equations''
q-alg/9701008}
\lref\curtjuan{C.~G.~Callan, Jr., J.~M.~Maldacena, \np{513}{1998}{198-212},
hep-th/9708147}
\lref\bak{D.~Bak, \pl{471}{1999}{149-154}, hep-th/9910135}
\lref\moriyama{S.~Moriyama,
hep-th/0003231}
\lref\wadia{A.~Dhar, G.~Mandal and S.~R.~Wadia, 
\mpl{A7}{1992}{3129-3146}\semi
A.~Dhar, G.~Mandal and S.~R.~Wadia, \ijmp{A8}{1993}{3811-3828}\semi
A.~Dhar, G.~Mandal and S.~R.~Wadia,  \mpl{A8}{1993}{3557-3568}\semi
A.~Dhar, G.~Mandal and S.~R.~Wadia,  \pl{329}{1994}{15-26}}

\lref\mateos{D.~Mateos, ``Noncommutative vs. commutative descriptions of
D-brane BIons'', hep-th/0002020}
\lref\mrs{S.~Minwala, M.~ van Raamsdonk, N.~Seiberg,
``Noncommutative Perturbative Dynamics'', hep-th/9912072}
\lref\nahm{W.~Nahm, \pl{90}{1980}{413}\semi
W.~Nahm, ``The Construction of All Self-Dual Multimonopoles
by the ADHM Method'', in ``Monopoles in quantum field theory'', Craigie et
al., Eds., World Scientific, Singapore (1982) \semi
N.J.~Hitchin, \cmp{89}{1983}{145}}
\lref\rs{M.~van Raamsdonk, N.~Seiberg, ``Comments of Noncommutative
Perturbative Dynamics'', hep-th/0002186, \jhep{0003}{2000}{035}}
\lref\k{M.~Kontsevich, ``Deformation quantization of Poisson
manifolds'', q-alg/9709040}
\lref\gms{R.~Gopakumar, S.~Minwala, A.~Strominger,
hep-th/0003160, \jhep{0005}{2000}{020}}
\lref\sst{N.~Seiberg, L.~Susskind, N.~Toumbas, hep-th/0005040}
\lref\sdual{R.~Gopakumar, S.~Minwala, J.~Maldacena, A.~Strominger,
hep-th/0005048\semi
O.~Ganor, G.~Rajesh, S.~Sethi, hep-th/00050046}
\lref\filk{T.~Filk, ``Divergencies in a Field Theory on  Quantum Space'',
\pl{376}{1996}{53}}
\lref\cf{A.~Cattaneo, G.~Felder, ``A Path Integral Approach to the Konstevich
Quantization Formula'', math.QA/9902090}

\lref\cds{A.~Connes, M.~Douglas, A.~Schwarz, \jhep{9802}{1998}{003}}
\lref\wtnc{E.~Witten, \np{268}{1986}{253}}
\lref\volker{V.~Schomerus, ``D-Branes and Deformation Quantization'',
\jhep{9906}{1999}{030}}
\lref\cg{E.~Corrigan, P.~Goddard, ``Construction of instanton and
monopole solutions and reciprocity'', \ap{154}{1984}{253}}

\lref\donaldson{S.K.~Donaldson, ``Instantons and Geometric
Invariant Theory", \cmp{93}{1984}{453-460}}

\lref\nakajima{H.~Nakajima, ``Lectures on Hilbert Schemes of
Points on Surfaces''\semi AMS University Lecture Series, 1999,
ISBN 0-8218-1956-9. }

\lref\neksch{N.~Nekrasov, A.~S.~Schwarz, hep-th/9802068,
\cmp{198}{1998}{689}}

\lref\freck{A.~Losev, N.~Nekrasov, S.~Shatashvili, ``The Freckled
Instantons'', {\tt hep-th/9908204}, Y.~Golfand Memorial Volume,
M.~Shifman Eds., World Scientific, Singapore, in press}

\lref\rkh{N.J.~Hitchin, A.~Karlhede, U.~Lindstrom, and M.~Rocek,
\cmp{108}{1987}{535}}

\lref\branek{H.~Braden, N.~Nekrasov, hep-th/9912019\semi
K.~Furuuchi, hep-th/9912047}

\lref\wilson{G.~ Wilson, ``Collisions of Calogero-Moser particles
and adelic Grassmannian", Invent. Math. 133 (1998) 1-41.}

\lref\gkp{S.~Gukov, I.~Klebanov, A.~Polyakov,
hep-th/9711112, \pl{423}{1998}{64-70}}
\lref\abs{O.~Aharony, M.~Berkooz, N.~Seiberg,
hep-th/9712117, \atmp{2}{1998}{119-153}}

\lref\abkss{O.~Aharony, M.~Berkooz, S.~Kachru, N.~Seiberg,
E.~Silverstein, hep-th/9707079, \atmp{1}{1998}{148-157}}

\lref\witsei{N.~Seiberg, E.~Witten, hep-th/9908142, \jhep{9909}{1999}{032}}
\lref\kinks{E.~Teo, C.~Ting, ``Monopoles, vortices and kinks in the
framework of noncommutative geometry'',
\pr{D56}{1997}{2291-2302}, hep-th/9706101}

\lref\manuel{D.-E.~Diaconescu, \np{503}{1997}{220-238}, hep-th/9608163}

\lref\genmnp{L.~Jiang,
``Dirac Monopole in Non-Commutative Space'', hep-th/0001073}
\lref\hashimoto{K.~Hashimoto, H.~Hata, S.~Moriyama,
hep-th/9910196, \jhep{9912}{1999}{021}\semi
A.~Hashimoto,
K.~Hashimoto, hep-th/9909202, \jhep{9911}{1999}{005}\semi
K.~Hashimoto, T.~Hirayama, hep-th/0002090}
\lref\hklm{J.~Harvey, P.~Kraus, F.~Larsen, E.~Martinec,
hep-th/0005031}

\lref\snyder{H.~S.~Snyder, ``Quantized Space-Time'', \pr{71}{1947}{38};
``The Electromagnetic Field in Quantized Space-Time'', \pr{72}{1947}{68}}
\lref\connes{A.~Connes, ``Noncommutative geometry'', Academic Press (1994)}

\lref\barsminic{I.~Bars, D.~Minic,
``Non-Commutative Geometry on a Discrete Periodic Lattice and Gauge Theory'',
hep-th/9910091}

\Title{\vbox{\baselineskip 10pt \hbox{PUPT-1932} \hbox{ITEP-TH-23/00}
\hbox{NSF-ITP-00-43} \hbox{hep-th/0005204} {\hbox{   }}}} {\vbox{\vskip
-30 true pt \centerline{MONOPOLES AND STRINGS IN}
\smallskip
\smallskip
  \centerline{NONCOMMUTATIVE GAUGE THEORY}
\medskip
\vskip4pt }} \vskip -20 true pt \centerline{ David J.~Gross
$^{1}$, Nikita A.~Nekrasov $^{2}$}
\smallskip\smallskip
\centerline{ $^{1}$ \it Institute for Theoretical Physics, University
of California Santa Barbara CA 93106}
\centerline{$^{2}$ \it Institute for Theoretical and Experimental
Physics, 117259 Moscow, Russia}
\centerline{$^{2}$ \it Joseph Henry
Laboratories, Princeton University, Princeton, New Jersey 08544}

\bigskip
\centerline{\rm e-mail: gross@itp.ucsb.edu,
nikita@feynman.princeton.edu}
\bigskip
\centerline{\bf Abstract} \vskip 1cm

We study some non-perturbative aspects of noncommutative gauge
theories.  We find  analytic solutions of the equations of motion,
for noncommutative U(1) gauge theory,
that describe magnetic monopoles with a finite tension string attached.
These solutions are non-singular, finite
and sourceless.
We identify the string with the
projection of a D-string ending
on a D3-brane in the presence of a constant $B$-field.

\Date{05/00}

\vfill\eject
\newsec{Introduction}

Recently there has been a revival of interest in field
theories on noncommutative spaces \snyder\connes, especially  those
  that emerge as various limits
of M theory compactifications
\cds. The latest circumstances in which
such theories were found involve D-branes in the presence of a
background Neveu-Schwarz $B$-field \volker\witsei.
The interest in such theories
is motivated by many analogies
between   noncommutative gauge theories and  large $N$
ordinary non-abelian gauge theories \filk\mrs, and also by  the
many features that noncommutative field theories share  with open string
theory \wtnc\mrs\rs.

In this paper we  study  some non-perturbative
dynamical objects in   noncommutative gauge theory,
specifically  four dimensional
gauge theory with an
adjoint Higgs field. The theory depends on a dimensionfull parameter
${\t}$ which enters the commutation relation between the coordinates of
the space:
$[x , x] \sim i {\t}$. We treat  only the bosonic  fields, but these
could  be a part of a supersymmetric multiplet, with ${\CN}=2$
supersymmetry or higher.
Such field theories arise on the world volume of D3-branes in the
presence of a background constant $B$-field along the D3-brane.

A D3-brane can be surrounded by other branes as well. For
example, in the Euclidean setup, a D-instanton could approach
the D3-brane. In fact, unless the D-instanton is dissolved inside
   the brane the combined system breaks supersymmetry \witsei. The
D3-D(-1) system can be rather simply described in terms of  a
noncommutative $U(1)$ gauge theory - the latter has
instanton-like  solutions \neksch. However, it turns out that the
``topology'' of the combined system is non-trivial (despite the
fact that the notion of a ``point'' on a noncommutative space makes very little
sense, the non-triviality of topology can be detected),
and it is this topology that supports the instantons \branek.

Another, perhaps even more interesting situation, is that
of a D-string that
ends on a  D3-brane. The endpoint of the D-string is a magnetic
charge for the gauge field on the D3-brane. In the commutative
case, in the absence of the $B$-field, the D-string is a straight
line, orthogonal to the D3-brane. It projects onto the
D3-brane in the form of a singular source, located at the point
where the D-string touches the  D3-brane. From the point of
view of the D3-brane   this is
a Dirac monopole, with   energy density that  diverges
at the origin.

The situation changes drastically when the $B$-field is turned on.
One can trade a constant background $B$-field with spatial
components for a constant background magnetic field. The latter pulls
the magnetic monopoles with the constant force. As a consequence, the
D-string bends \hashimoto, in order for its tension
to compensate the magnetic force.
It projects to the D3-brane as a half-line with finite
tension. It is quite fascinating to see, as we shall explicitly verify,
that the shadow of this string is seen by the
noncommutative gauge theory. The $U(1)$ noncommutative gauge
theory has a
monopole solution, that is everywhere non-singular, and whose
energy density localizes along a half-line, making up a semi-infinite string.
We should stress that the non-singularity of the solution is the
non-perturbative in ${\t}$ property, it
couldn't be seen by the expansion in ${\t}$ around the Dirac monopole
\genmnp.

The fact that all the fields involved are non-singular, and that
the solution is in fact a solution to the noncommutative version
of the Bogomolny equations everywhere,
makes us suspect that the string in the monopole solution is an
intrinsic object of the gauge theory. As such, one could expect
that the noncommutative gauge theory {\it holographically}
describes strings as well. This statement is further supported by the
fact that in the limit of very large $B$-field (the limit
which must be well described by the non-commutative gauge theory \witsei)
the D-string almost lies on top of the D3-brane, practically
dissolving in it.

Finally, by applying S-duality \sdual\
one could map the solution we found into the
solution describing the electric flux tube, represented by the
fundamental string \hklm.
In this way one may hope to arrive at the description of the confining strings
in the noncommutatve Yang-Mills theories.
Notice however, that the
S-duality maps the theory with the spatial noncommutativity to that of the
temporal noncommutativity, with all its surprises \gkp\sst, in addition to
the strong coupling \sdual.

The outline of this paper is as follows. In Section 2 we consider some general
features of noncommutative
field theory, and discuss how it is convenient to work in the  Fock space
in which the coordinates are expresses as creation and annihilation operators.
In Section 3 we  construct the Green's function of the Laplace operator on
noncommutative
spaces, which illustrates the smearing of space induced by the
noncommutativity of the coordinates.
We also give a brief introduction to noncommuative gauge theories.

In  Section 4  we  set up the equations for BPS solutions
of four dimension noncommutative gauge theories.  We review Nahm's
construction of
commutative monopoles,  exhibit the SU(2) monopole, as well as the Dirac
monopole in this framework.
Section 5 is devoted  to the construction of the explicit solution of the
BPS equations for the
$U(1)$ noncommutative gauge theory coupled to a scalar field. The properties
of the solution are analyzed in Section 6.
We conclude, in Section 7, with a discussion of the implications of our
analysis.

\vfill\eject
\ndt {\bf Acknowledgements.}

\ndt{}We would like to thank  D.~Bak, S.~Cherkis,
D.~-E.~Diaconescu, S.~Giddings, A.~Hashimoto, K.~Hashimoto,
N.~Itzhaki, I.~Klebanov, T.~Piatina, A.~Polyakov, A.~Schwarz,
S.~Shatashvili, and K.~Selivanov for discussions. Our research was
partially supported by NSF under the grant PHY94-07194, in
addition, research of NN was supported by Robert H.~Dicke
fellowship from Princeton University, partly by RFFI under grant
00-02-16530, partly by the grant 00-15-96557 for scientific
schools. NN is grateful to ITP, UC Santa Barbara, CIT-USC Center,
and CGTP at Duke University for their hospitality during various
stages of this work.

\newsec{Noncommutative Field Theory}

Consider   space-time  with coordinates  $x^i$,
$i=1, \ldots, d$ which obey the following commutation relations:
\eqn\cmrl{[x^i, x^j] = i {\t}^{ij}\ , } where ${\t}^{ij}$ is a
constant asymmetric matrix of  rank $2r \leq d$. By
noncommutative space-time one means the algebra ${\CA}_{\t}$
generated by the $x^i$ satisfying \cmrl , together  with some extra
conditions on
the allowed expressions of the  $x^i$. The elements of ${\CA}_{\t}$ can
be identified with   ordinary functions on ${\bf R}^d$, with the
product of two functions $f$ and $g$ given by the Moyal formula (or
star product):
\eqn\myl{f \star g \, (x)= {\exp} \left[ {i \over 2} {\t}^{ij}
{{\p}\over{\p x_{1}^{i}}} {{\p}\over{\p x_{2}^{j}}}
\right] f (x_{1}) g (x_{2}) \vert_{x_{1} = x_{2} = x}\ .}

\ndt A field theory is defined as usual  by constructing an action,
say in the case of a scalar
field theory,
$${\CL} ({\Phi}) = \int d^{d}x \,\, \left[ {\p}_i {\Phi}
\star {\p}_i {\Phi} + V ({\Phi}) + \ldots \right]. $$ The symbol $\int
d^{d}x$ is a notation for a trace, ${\Tr}$, on the algebra
${\CA}_{\t}$. When one works on  compact noncommutative
manifolds (compact manifolds, whose algebra of functions is
deformed, e.g. by the techniques of \k), for example the
noncommutative torus,  then the  trace is the usual trace, i.e. the
linear map ${\CA}_{\t} \to {\IC}$, such that ${\Tr} [ a, b] = 0$.
On  ${\bf R}^d$ the notion of the trace is trickier, in
particular the trace of the commutator may not vanish, just as
the integral of a total derivative may not vanish. We shall
encounter such effects in our discussion below, so instead of
giving formal definitions at this point  we shall treat explicit examples
later.

The Lagrangian of a field theory involves derivatives. The
derivative ${\p}_i$ is the infinitesimal automorphism of the
algebra \cmrl: \eqn\auto{x^i \mapsto x^i + {\ve}^i,} where
${\ve}^i$ is a $c$-number. For the algebra \cmrl\ this automorphism
is internal:
\eqn\intrn{{\p}_i {\Psi} = i  {\t}_{ij} [ {\Psi}, x^j],}
where ${\t}_{ij}$ is the inverse of ${\t}^{ij}$, namely
${\t}_{ij}{\t}^{jk}=\delta^k_i$.
  In contrast,  on the torus generated by $U_l = {\exp} 2\pi i x^l$, it is an outer automorphism. This difference is  crucial in the
analysis of noncommutative  gauge theories.

By an orthogonal change of coordinates we can map the Poisson
tensor ${\t}_{ij}$ onto its canonical form: $$ x^i \mapsto z_a,
{\zb}_{a}, \quad a = 1, \ldots, r\ ; \quad y_{b}, \quad b = 1,
\ldots, d- 2r ,$$ so that:
\eqn\ncm{\eqalign{&[y_a,y_b]= [y_b, z_a] =
[y_b, {\zb_a}] =
0, \quad [z_{a}, {\zb}_{b}] = 2{\t}_{a}{\d}_{ab} , {\quad} {\t}_{a}
> 0 \cr & \quad ds^2 =dx^2_i+ dy_b^2 = dz_a d{\zb}_{a} + dy_b^2 .\cr} }

Since $z (\zb)$ satisfy (up to a constant) the commutation relations
of creation (annihilation) operators we can
identify functions $f(x,y)$ with   operator valued functions of the $y_a$ in
the Fock space of the
$r$ creation and annihilation operators (the operators in the Fock space
are widely used in the studies of noncommutative theories and matrix models,
for their applications to the latter see \wadia):
\eqn\fock{ {\a}_a=z_a/\sqrt{2\t_a},\quad
{\a}^{\dagger}_a=\zb_a/\sqrt{2\t_a},\quad [{\a}_{a}, \a^{\dagger}_b] =\d_{ab}.}
Since we shall be dealing with scale invariant theories in which the
only scales will be the $\t_a$ we shall
set all $2\t_a=1$. When desired, the $\t_a$'s can be introduced by
rescaling the coordinates,
$z_a \to z_a/\sqrt{2\t_a}$. Let ${\hat n}_{a} = {\a}^{\dagger}_{a}{\a}_{a}$
be the $a$'th number operator.

The procedure that maps
  ordinary commutative functions onto   operators in the Fock
space acted on by $z_a, {\zb}_a$ is called Weyl ordering and is
defined by: \eqn\wlor{f\left(x= \left( Z_a, {\bar Z}_a \right)\right)
\mapsto {\hat f(z_a,
{\bar z}_a)}
=  \int f(x) \, {{d^{2r} x \,\, d^{2r} p }\over{(2{\pi})^{2r}}} \,
  \,  e^{ i \left( {\bar p}_a \left(  z_a  - Z_a \right) + p_{a}
\left( {\zb}_a -
  {\bar Z}_a \right) \right)}.}
It is easy to see that \eqn\product{  {\rm if } \quad f \mapsto \hat
f, \quad g \mapsto \hat g \quad {\rm then }\quad f\star g \mapsto
\hat f \hat g .}

\ndt A useful formula is for the matrix elements of $\hat f$ in the
coherent state basis
\eqn\coherent{\langle {\xi} \vert {\hat f} \vert {\eta} \rangle =
\int f\left(Z, {\bar Z} \right) {{d^{r} Z \, d^{r} {\bar
Z}}\over{(2{\pi}i)^{2r}}}
\, e^{\xi \cdot \eta - 2 ( {\xi} - {\bar Z} ) \cdot ( {\eta} - Z )}}
where $\langle {\xi} \vert$ and $\vert {\eta} \rangle$
  are coherent states:
$\langle {\xi} \vert  = \langle {\bf 0} \vert
{\exp} \left( {\xi}_a z_a \right) , \qquad \vert {\eta} \rangle =
{\exp} \left( {\eta}_a {\zb}_a \right) \vert {\bf 0} \rangle .
$
\ndt From \coherent\ we can extract the matrix
elements of ${\hat f}$ between the standard oscillator states
  by:
  \eqn\mtrx{\langle {\bf
k} \vert {\hat f} \vert {\bf l} \rangle = {1\over{\sqrt{{\bf
k}!{\bf l}!}}} {\p}^{\bf k}_{\xi} {\p}^{\bf l}_{\eta}\vert_{\xi = \eta = 0}
\langle {\xi} \vert {\hat f} \vert {\eta} \rangle, } where $\bf k$,
$\bf l$ are the vectors of the occupation numbers, e.g. ${\bf k }
= (k_1, \ldots, k_{r})$.

Given the operator $\hat f$, or its matrix elements, in the
coherent state or occupation number basis, one can easily
reconstruct the function $f(x)$ to which it corresponds. For
example, consider  the simplest case where $r=1, \ d=3$, that will
be our interest below. Furthermore, consider functions that are
axially symmetric. This means that $f(x)=f(r,x_3)$, where
$r=\sqrt{x_1^2+x_2^2}$; or equivalently that $\langle k \vert \hat
f(x_3)\vert l\rangle= \d_{kl} f_{l}(x_3)$. Then  to reconstruct
the function $f(x)$, from the matrix elements $ f_{l}(x_3)$ one
uses: \eqn\recon{  f(r,x_3)= 2 \sum_{l=0}^\infty (-1)^{l}
f_{l}(x_3) L_l (4r^2) e^{-2r^2} \ ,} where $L_l (4r^2)$ are
Laguerre polynomials.

\newsec{Scalar Field Green's functions}

An interesting property of noncommutative field theory is its
similarity with lattice field theory, namely the noncommutativity of the
coordinates introduces a smearing of space.  We shall
illustrate this similarity by examining the
Green's functions of the Laplace operator on  noncommutative
space-time.

\subsec{Sources}

Consider the noncommutative version of the
equation for the Green's function
  ${\Delta}_{x} G(x, x^{\prime}) = {\d} (x - x^{\prime})$.
Recalling \intrn, the Laplace
operator can be  rewritten as follows:
\eqn\lplc{{\hat \Delta} = {{\p^2}\over{{\p} y_{b}^2}} - 4 {\t}_{a}^{-2}
[ [ \cdot, z_{a}], {\zb}_{a}]. }
Thus the noncommutative
equation for the Green's function is
\eqn\gf{ {{\p^2}\over{{\p} y_{b}^2}} \hat G(z,\zb, y; z^{\prime},
\zb^{\prime}, y^{\prime})-4[
[ \hat G(z,\bar z,y;z^{\prime},\bar z^{\prime}, y^{\prime}), z_{a}],
{\zb}_{a}]=\hat \delta(z,\bar z;z^{\prime},\zb^{\prime}) {\d} (y-y^{\prime}),}
where we have introduced  two copies of the algebra ${\CA}_{\t}$,
generated by $z_a, {\zb}_a, z_a^{\prime}, {\zb}_a^{\prime}$ and  $\hat
\delta, \hat G$
are operators in the tensor product of
two Fock spaces
$$
{\CH}_{1,2}  = {\CH}_{1} \otimes {\CH}_{2}
$$
spanned by $\vert {\bf l}_1, {\bf l}_2 \rangle = \vert {\bf l}_1
\rangle \otimes \vert
{\bf l}_{2} \rangle$.

The expression for the  delta function,
${\hat \d} (z,\zb; z^{\prime} ,\zb^{\prime})$,  is now easy to obtain directly
in the  coherent state basis , using the (tensor product form of) \coherent.
In terms of $\vert {\eta}_1, {\eta}_2 \rangle =
\vert {\eta}_1 \rangle \otimes \vert {\eta}_2 \rangle$
we have
\eqn\dltii{\eqalign{ \langle {\xi}_{1}, \xi_{2} \vert {\hat\d} \vert
{\eta}_{1}, \eta_{2} \rangle =
\int &  e^{{\xi}_1 \cdot {\eta}_1 + {\xi}_2 \cdot {\eta}_2  -
2 ( {\xi}_1 - {\bar Z} ) \cdot ( {\eta}_1 - Z ) -
2 ( {\xi}_2 - {\bar Z} ) \cdot ( {\eta}_2 - Z ) } d^r Z d^r {\bar Z}   \cr =
& e^{{\xi}_1 \cdot {\eta}_2 + {\xi}_2 \cdot {\eta}_1}. \cr}}

The matrix elements of $\d $ in the occupation number basis are:
\eqn\dltiv{\langle {\bf k}_{1,2} \vert {\hat \d} \vert {\bf l}_{1,2} \rangle =
{\d}_{{\bf k}_1, {\bf l}_2} {\d}_{{\bf k}_2, {\bf l}_1}.}
Thus ${\hat \d}$ is a permutation operator $P: {\CH} \otimes {\CH} \to
   {\CH} \otimes {\CH}$, $P (e_1 \otimes e_2 ) =  e_2 \otimes e_1$, and
squares to the identity operator $P^2 = Id$. It is easy to verify that
$\hat \d$ satisfies the defining property of the delta-function, namely
\eqn\trace{ {\Tr}_x^{\prime}\left[ \hat \d (x,x^{\prime}) \hat
f(x^{\prime})\right] = \hat f(x)}

What is the noncommutative  version of a source localized at the origin,
namely $\delta^{2r}(x)$?  Using \coherent\ we see that
\eqn\dltt{\langle {\xi} \vert {\hat \d} \vert {\eta} \rangle =
\int \delta^2(Z) {{d^{r} Z \, d^{r} {\bar Z}}\over{(2{\pi}i)^{2r}}}
\, e^{\xi \cdot \eta - 2 ( {\xi} - {\bar Z} ) \cdot ( {\eta} - Z
)}=e^{-\xi \cdot \eta},}
or in the occupation number basis:
\eqn\dltor{\langle {\bf k} \vert {\hat \d} \vert {\bf l} \rangle =
  {\d}_{{\bf k}, {\bf l}} (-1)^{\vert {\bf k} \vert}, \quad  \hat \d
= (-1)^{\hat {\bf n}}} with
$\vert {\bf k} \vert = \sum_a k_a,  \hat {\bf n} =\sum_a \hat n_a $.
In this way the delta
function becomes an operator in the Fock space with the spectrum
of the form of the diffraction rings. Note that $(\hat \delta(x))^2 =Id$,
which is the transform of the constant function.

Alternatively we can relate
\dltiv\  to  \dltor.
by  passing to the center-of-mass frame:
\eqn\com{z^c = {1\over{\sqrt{2}}} \left( z + z^{\prime} \right), z^r =
{1\over{\sqrt{2}}}
\left( z - z^{\prime} \right)} and similarly for ${\zb}, {\zb}^{\prime}$.
The expression \dltor\
is written in the number basis for the
operators $z^r, {\zb}^r$.
The transformation \com\ is a unitary one:
\eqn\uni{\eqalign{S z^r S^{\dagger} = z, & \qquad
S z^c S^{\dagger} = z^{\prime}, \cr
S z S^{\dagger} = z^{c}, & \qquad S z^{\prime} S^{\dagger}  = - z^r \cr
S = {\exp} {{\pi}\over 4}  & \left( {\zb}^{\prime} \cdot z -
{\zb} \cdot z^{\prime} \right) \cr}}
It is easy to check that $S P = P S^{\dagger}$. Therefore
$S P S^{\dagger} = S^2 P$.
Now, consider $U = S^2$. It acts as follows:
\eqn\rot{U z U^{\dagger} = z^{\prime}, \quad U z^{\prime} U^{\dagger} = - z}
Let us now apply the $S$ transformation to the delta function:
$$
SP S^{\dagger} \vert {\bf l}_{1} \rangle \otimes \vert {\bf l}_2 \rangle
= U \vert {\bf l}_2 \rangle \otimes \vert {\bf l}_{1} \rangle =
(-1)^{\vert {\bf l}_1 \vert} \vert {\bf l}_1 \rangle \otimes
\vert {\bf l}_2 \rangle
$$
i.e. we get complete agreement with \dltor.

{}Thus in the noncommutative case
we cannot construct a truly localized source. The transform of
$\delta^{2r}(x)$, in which
the noncommuting coordinates are all localized at the origin, is
spread out over all of space.
The most localized source we can construct in the noncommutative
case is a Gaussian wave packet
$D(x)= \exp(- 2 Z{\bar Z})$, whose transform is
\eqn\gauss{ \hat D = \vert {\bf 0} \rangle \langle  {\bf 0}\vert ,
\quad \langle {\xi} \vert {\hat D} \vert {\eta} \rangle = 1,\quad
\langle {\bf k} \vert {\hat D} \vert {\bf l} \rangle =\delta_{{\bf
k},{\bf l}}\delta_{{\bf l},{\bf 0}}}

One can also develop the similar analysis for finite lattices,
in which case one gets the finite matrix versions of the operators
\dltor\dltiv (see \barsminic).

\subsec{Green's functions}

We now consider  the Laplace equation for the Green's function, $\hat G$.
Consider a
function (an element of ${\CA}_{\t}$) that commutes with all
$N_{a}$'s. In the commutative language this means that the
functions we wish to look at are invariant under rotations of the
$z_{a}, {\zb}_{a}$ two-planes. We take  $\hat G$ to be such a
function.  On such functions the Laplacian
acts as follows:
\eqn\lplci{{\Delta}G_{\bf n} = {{\p^2}\over{{\p}
y_{b}^2}} G_{\bf n} + 4\sum_{a}\left(  -(2n_a +1) G_{\bf n} + (n_a +
1)G_{{\bf n + e}_{a}} + n_{a} G_{{\bf n - e}_{a}} \right)} where
\eqn\ns{{\bf n} =
\left( n_1, \ldots, n_r \right), \quad {\bf e}_{a} = \left( 0, 0, \ldots,
1_{\lower.12in\hbox{\kern -.06in ${\scriptstyle{\widehat a}}$}} \, ,
\ldots, 0\right), \quad  \langle  {\bf n}
\vert
\hat G\vert {\bf n}^{\prime} \rangle =\delta_{{\bf n}^{\prime},{\bf
n}}G_{\bf n}.}
The formula \lplci\
requires the following comment: when evaluating the right hand side
of \lplci\ the
number operators $n_{a}$ must be evaluated first, so that if some
of the $n_a$'s vanish the whole expression $n_{a}G_{{\bf n - e}_{a}}$
must be set to zero, no matter how singular the {\it analytic
expression} for $G_{{\bf n - e}_{a}}$ may look.

One can also rewrite the
Laplacian \lplci\ using the finite difference operators:
${\CD}_{a}, {\CN}_{a} $: \eqn\difrep{ {\CD}_{a} G_{\bf n} = G_{\bf n} -
G_{{\bf n - e}_{a}}, \quad {\CN}_{a} G_{\bf n} = (n_{a} + 1)
G_{{\bf n + e}_{a}}}
\eqn\lplcd{{\Delta} = {{\p^2}\over{{\p} y_{b}^2}} + 4 {\CD}_{a}
{\CN}_{a} {\CD}_{a}\ .} The operators ${\CN}, {\CD}$ form  a
Heisenberg algebra:
\eqn\difalg{[ {\CD}_{a}, {\CN}_{b} ] =
{\d}_{ab}.} {}Let us compare the expressions \lplci,\lplcd\ to
their commutative analogues. Let $y_{b}, Z_a, {\bar Z}_a$ denote
the coordinates on the commutative space-time with the metric $$
ds^2
=
dy_b^2 + {\half}dZ_a d {\bar Z}_a \ ,$$ and consider   functions
that depend only on  $y_b$
and $R_a = \vert Z_a \vert^2$. On such a function, say  ${\CG}$, the
Laplacian acts as follows: \eqn\clpl{{\Delta} {\CG} =
{{\p^2}\over{{\p} y_{b}^2}} {\CG} + 4 \sum_a {{\p}\over{\p R_a}}
R_{a} {{\p}\over{\p R_{a}}} {\CG}.} The operators $R_a$ and
${\p}_{R_{b}}$ form the same Heisenberg algebra \difalg as ${\CN},
{\CD}$. Consequently, by
mapping the representation  \difrep\ to the standard
representation of the Heisenberg algebra acting on  functions of $R_a$,
we can map the Green's function of the
noncommutative Laplacian to that of the commutative one.

{}Let us note, however, that the algebra \difalg\ is represented
by functions on the whole space ${\bf R}^{r}$, whereas  the
variables $R_{a}$, by definition, must be positive. The same
comment applies to the variables $n_a \geq 0$. This boundnessness
of the domain in the definition of these functions produces the source
terms in the Laplace (or other) equations that they obey.

{}Now   construct  the Laplace transform of the function ${\CG}$:
\eqn\lt{{\tilde \CG} (t) = \int_0^{\infty} \prod_a dR_a e^{-t_{a} R_{a}}
{\CG}(R_{a}).} {}At the same time  we construct the generating
function associated with $G_{\bf n}$:
\eqn\gf{{\hat G} (t)  = \sum_{\bf n} \prod_{a} ( 1 -
t_{a})^{n_{a}} G_{\bf n}. }
It is easy to see that the  Laplacian  operators acting on both $\tilde \CG(t)$
and on ${\hat G} (t)$,
are mapped to the same operator:
\eqn\trlp{{\hat\Delta} = {{\p}\over{{\p} y_{b}^2}}
  - 4 t_{a} {{\p}\over{{\p} t_{a}}} t_{a}, }
(as long as we assume  that the functions don't grow too fast at infinity or
at zero)
{} In this way we define a  map from   functions of continuous $R$ coordinates
to   functions of discrete $\bf n$: \eqn\mps{G_{\bf n} =
\int_{0}^{\infty} {\CG}_{R}  \prod_{a} {{R_a^{n_a}}\over{n_a !}}
dR_a e^{-R_a}.}
{} Note that as ${\bf n} \to \infty$ the saddle point approximation gives $$
G_{\bf n} \sim {\CG}_{R_a = n_a} \ .$$

We can use this to construct the noncommutative version of the
Green's function.
{} For example, take $d = 2r +1$,
then the $2r+1$ - dimensional Green's function
$${\CG} (y, R) = {1\over{\left( y^2 + \sum_a R_a \right)^{r +
{\half}}}},
$$
{}transforms into:
\eqn\thrd{ \quad G_{\bf n} = \int_{0}^{\infty} {\rm d} s {{s^{r -
{3\over 2}}}\over{(1 + s)^{\vert \bf n \vert +r}}} e^{-s y^2}, \quad \vert
\bf n \vert =
\sum_a n_a\ .} {}The noncommutative function is everywhere non-singular: $$
G_{0} (y) \sim - \sqrt{\pi} \left( 2 {{(r - {3\over 2})!}\over{(r
- 2)!}} + {\vert y \vert} + \ldots \right), \quad y \to \ . 0$$
Thus the  map renders   functions smoother at the
origin.

{\ndt}Another example is when $d = 2r$. Then  we have:
\eqn\frth{{\CG} (R) = {1\over{\left( \sum_a R_a \right)^{r-1}}},}
\eqn\even{G_{\bf n} = \int_0^1 {\rm d}{\l} \, {\l}^{r-2} (1 - {\l}
)^{\vert \bf n \vert} = {{(r-2)! \vert {\bf n} \vert !}\over{( {\vert {\bf
n} \vert} +r -1)!}} .} which is also non-singular
everywhere (for $r > 1$). In two dimensions ($r=1$) we get:
\eqn\twod{{\CG} (R) = {\rm log}\left( {\m} R \right) \Rightarrow
G_{n} = {\rm log}{\m} + {\psi} (n+1) = {\rm log} {\m} - C +
\sum_{k=1}^{N} {1\over{k}} .} The formula \even\ is  not applicable
here since it gives a logarithmically divergent integral, of purely infrared
origin.
However, the divergence is $n$ independent, so that it affects $G_n$
by an additive
constant, i.e. by a zero mode of the Laplacian. The
presence of the divergence is reflected in the fact that the
cutoff, ${\rm log}{\m}$, in the commutative Green's function
appears in the noncommutative formula \twod.

{}The commutative Green's function
${\CG} (R)$ solved   Laplace's equation with a delta function  source:
$${\Delta} {\CG(R)} = {\delta}^{d}(x)$$ Let us see what the source is
equal to now. For simplicity let us work in even number of
dimensions, $G_{\bf n} = G ( {\vert {\bf n} \vert})$:
\eqn\evlp{{1\over 4} {\Delta}G (n)  = ( n
 + r) G (n+1)
+ n G (n-1)
-
(2n + r) G (n) = -  \int_{0}^{1} {\rm d} \left( {\l}^{r} ( 1-
{\l})^n \right) = - {\d}_{n,0}.}

{\ndt}So the source got smoothed out: \eqn\srce{{\delta}^{d}(x) =
\prod {\delta} (R_a) \prod {\delta} (y_b) \mapsto \prod
{\delta}_{N_{a}, 0} \prod {\delta} (y_b).}
  In this way the
noncommutativity of the space-time looks similar to the lattice
regularization (although in the spherical rather then Cartesian
way). However, by the above analysis, the formula \evlp\ means that
we have ended up with the Gaussian source $D$
as in \gauss.

On the other hand, the
solution to the equation $$ {\Delta} {\hat G} = {\hat \d}$$ with the
localized delta function source is
also easy to produce: one simply applies the map \wlor\ to the ordinary
Green's function ${\CG}_{R}$. It is amusing that the result is
close to the formula \even, namely the Green's function is
again diagonal in the eigenbasis of the occupation number
operators $n_a$ and depends only on $n = \vert {\bf n} \vert$:
\eqn\eveni{{\hat G}_{\bf n} = {\hat G} (n)  =
\int_{0}^{2} {\rm d}{\l} \, {\l}^{r-2} (
1 - {\l} )^{n}\ . }
{}However, \eveni\ and \even\ differ considerably. In four dimensions,
$r=2$,  the difference is
striking:
\eqn\frdm{G (n) = {1\over{n+1}}, \quad {\hat G} (n) =
{{1 + (-1)^{n}}\over{n+1}}.}
What is also striking is the failure of
the classical limit for $r > 2$: one might expect that, for large
$n$, the Green's function ${\hat G} (n)$ would go over to its classical
counterpart ${\CG}_{R} \sim {1\over{n^{r-1}}}$. The integral
\even\ indeed has this property - the integrand is peaked at ${\l}
= 0$ and the saddle point gives precisely the expected
asymptotics. But the integral \eveni\ has another saddle point at
${\l} =2$ which yields the leading asymptotics for $r > 2$ $$ {\hat
G} (n) \sim {{(-1)^{n} \, 2^{r-2}}\over{n+1}}, \qquad n \to \infty .
$$
{}The lesson to be drawn from here is that  highly
localized distributions (the delta function is such a
distribution) become the operators spread out  over all the Fock space
that they  act in, while the operators whose range is
comparatively small (such as the Gaussian) correspond in fact to the
distributions
with  finite support of order  $\sim \sqrt{\t}$.

{}Finally, to construct the Green's function  ${\hat G}(x - x^{\prime})$
we have to use
the formula:
$$
{\hat G}( x - x^{\prime}) = S^{\dagger} \left( {\hat G} (x) {\otimes}
{\Id}_{x^{\prime}}
\right) S\ ,
$$
which gives (for $d=2r$):
\eqn\grtwo{G (x - x^{\prime}) =
\int_0^2 {\rm d} {\l} \, {\l}^{r-2} (1 - {\l})^{{\half} (z - z^{\prime})
\cdot ( {\zb} - {\zb}^{\prime} )}\ . }

\subsec{Gauge theory on  noncommutative space}

{}In an  ordinary gauge theory with gauge group $G$ the gauge fields
are connections in some principal $G$-bundle. The matter
fields are the sections of the vector bundles with the structure
group $G$. Noncommutative vector bundles are defined as
projective modules over the algebra ${\CA}_{\t}$. This definition
captures the following two properties of  ordinary vector
bundles: i) the sections of the bundle can be multiplied by
functions on the base manifold - in this way the space of sections
is acted on (linearly) by the space of functions --- the
definition of a module; ii) every vector bundle can be made
trivial by the appropriate addition of another vector bundle - this is the
definition of the projective module - it becomes free (equals to a
direct sum of several copies of the algebra ${\CA}_{\t}$) when we
add  another module.

{}Now suppose we are given a module $M$ over the algebra
${\CA}_{\t}$. In the noncommutative case there are two types of
modules, left  and right. The elements $m_{\bf l}$ of the left
module are multiplied by the elements $a$ of the algebra from the
left, while the elements of the right module are multiplied from
the right: $$ a: m_{\bf l} \mapsto a m_{\bf l}, \quad m_{\bf r}
\mapsto m_{\bf r} a\ . $$ The left module over an algebra ${\CA}$ is
a right module over the algebra ${\CA}^{\prime}$ which is
obtained from ${\CA}$ by reversing the order of multiplication: $$
a \star^{\prime} b = b \star a\ .$$ The notion of the left/right
modules is analogous to the notion of   chiral matter fields.

{}The connection ${\nabla}$ is the operator $$ {\nabla} : {\bf R}^d
\times M \to M, \quad {\nabla}_{\ve} (m) \in M, \qquad {\ve} \in
{\bf R}^d, \, m \in M \ ,  $$ where ${\bf R}^d$ denotes the commutative
vector space, the Lie algebra of the automorphism group generated
by \auto. The  connection is required to obey the Leibnitz rule:
\eqn\leibn{\eqalign{& {\nabla}_{\ve} ( a m_{\bf l} ) = {\ve}^i (
{\p}_i a ) m_{\bf l} + a {\nabla}_{\ve} m_{\bf l} \cr &
{\nabla}_{\ve} ( m_{\bf r} a ) = m_{\bf r}  {\ve}^i ( {\p}_i a ) +
( {\nabla}_{\ve} m_{\bf r} ) a \ . \cr}}

{}As usual, one defines the curvature $F_{ij} = [{\nabla}_i, {\nabla}_j]$ -
the operator
${\Lambda}^2 {\bf R}^d \times M \to M$ which commutes with
the multiplication by $a \in {\CA}_{\t}$. In other words, $F_{ij} \in {\rm
End}_{\CA}(M)$.
If the right (left) module $M$ is free, i.e. it is a sum of several copies
of the algebra
${\CA}$ itself, then the connection ${\nabla}_i$ can be written as
$$
{\nabla}_i = {\p}_i + A_i
$$
where $A_i$ is the operator of the left (right) multiplication
by the matrix with ${\CA}$-valued entries. In the same operator sense the
curvature obeys the standard identity:
$$
F_{ij} = {\p}_i A_j - {\p}_j A_i + A_i A_j - A_j A_i \ .
$$
{}Given the module $M$ one can multiply it by a free module
${\CA}^{\oplus N}$ to make it a module over an algebra ${\rm
Mat}_{N \times N}({\CA})$ of matrices with elements from $\CA$. In
the non-abelian gauge theory over ${\CA}_{\t}$ we are interested
in projective  modules over ${\rm Mat}_{N \times N}({\CA}_{\t})$.
If the algebra ${\CA}_{\t}$ (or perhaps its subalgebra) has a trace,
${\Tr}$, then the algebra ${\rm Mat}_{N \times N} ({\CA}_{\t})$ has
a trace given by the composition of a usual matrix trace and
${\Tr}$.

{}It is a peculiar property of the noncommutative algebras that
the algebras ${\CA}$
and ${\rm Mat}_{N \times N}({\CA})$ have much  in common. These
algebras are called Morita equivalent and under some additional
conditions the gauge theories over ${\CA}$ and over ${\rm
Mat}_{N}({\CA})$ are also equivalent. This phenomenon is
responsible for the similarity between the ``abelian
noncommutative'' and ``non-abelian commutative'' theories.

\newsec{Monopoles and Instantons}

\subsec{Lagrangian and couplings}

{}After the preparations of the previous section
the Lagrangian for the gauge theory is
given  by: \eqn\glagr{{\CL}(A) = -{1\over{4g_{\rm YM}^2}}
\sum_{i,j} {\Tr} [{\nabla}_i, {\nabla}_j]^2 \ .} If   additional
charged matter fields are present (elements ${\Phi}$ of a module $M$)
then the Lagrangian becomes: \eqn\lagint{{\CL}(A, {\Phi}) = {\CL}
(A) + \sum_i {\Tr} {\nabla}_i {\Phi}  {\nabla}_i {\Phi} +
\ldots}
{}
The equations of motion following from the Lagrangian \glagr\ are:
\eqn\eqsm{\sum_i [{\nabla}_i, F_{ij}] = 0}
{}In four dimensions, $d=4$, the Euclidean version of \eqsm\ can be
solved by solving the first order instanton equations:
\eqn\inst{F_{ij} = \pm {1\over 2} {\e}_{ijkl} F_{kl}\ ,} as follows
from the Bianci identity, which holds irrespectively of the
commutativity: \eqn\bian{[{\nabla}_i, F_{kl} ] + [ {\nabla}_{k},
F_{li} ] + [ {\nabla}_{l}, F_{ik} ] = 0\ .}

{}Introduce the complex coordinates: $z_1 = x_1 + i x_2 = x_{+}$,
$z_2 = x_3 + i x_4$. The instanton equations read:
\eqn\inst{\eqalign{& \qquad\quad F_{z_1 z_2} \quad = \quad  0 \cr
& F_{z_1 {\zb}_1} + F_{z_2 {\zb}_2} \quad = \quad 0 \ . \cr}} The
first equation in \inst\ can be solved (locally) as follows:
\eqn\ynga{A_{{\zb}_a} = {\xi}^{-1} {\pb}_{{\zb}_a} {\xi}, \quad
A_{z_a} = - {\p}_{z_a} {\xi} {\xi}^{-1}\ .} with $\xi$   a
Hermitian matrix. Then the second equation in \inst\ becomes
Yang's equation: \eqn\ynge{\sum_{a=1}^2 {\pb}_{z_a} \left(
{\p}_{z_a} {\xi}^2 {\xi}^{-2} \right) = 0\ .} This ansatz works in
the noncommutative setup   as well \neksch.

{}If we look for the solutions to \inst, that are invariant under
translations in the $4$'th direction then we will find the
monopoles of the gauge theory with an adjoint scalar Higgs field,
where the role of the Higgs field is played by the component $A_4$
of the gauge field. The equations \inst\ in this case are called
the  Bogomolny equations, and they can be analyzed in the
commutative case via Nahm's ansatz \nahm.

{}The action \glagr\ becomes the energy functional for the coupled
gauge-adjoint Higgs system:
\eqn\enrg{{\CE} = {1\over{4g_{\rm YM}^2}} \int d^3 x \, \sqrt{{\rm det}G}
{\Tr} \left( G^{ii^{\prime}} G^{jj^{\prime}}
F_{ij} \star F_{i^{\prime}j^{\prime}} + G^{ij} {\nabla}_i {\Phi} \star
{\nabla}_j {\Phi} \right)}
where for the sake of generality we have introduced a constant
metric $G_{ij}$.

\sssec{\rm Open \quad and \quad closed \quad string \quad moduli.}

\ndt{} \enrg\ emerges in the decoupling limit of
a  D3-brane in  the Type IIB string theory in a background
with a constant Neveu-Schwarz B-field. Let us recall
 the relation of the parameters of the actions \glagr, \ \enrg\
and the string theory parameters, before  taking
the Seiberg-Witten limit \witsei.

We start with the D3-brane whose worldvolume is occupying the 0123
directions, and turn on a $B$-field: \eqn\bfi{ {\half} B dx^1
\wedge dx^2} The indices $i,j$ below will run from $1$ to $3$. We
assume that the closed string metric $g_{ij}$ is flat, and the
closed string coupling $g_{s}$ is small. According to \witsei\ the
gauge theory on the D3-brane is described by a  Lagrangian, which,
when  restricted to time-independent fields,  coincides with
\enrg, whose parameters $$ G_{ij}, {\t}^{ij}, g_{\rm YM}^2 \ , $$
are related to $$ g_{ij}, B_{ij}, g_{s} $$ as follows:
\eqn\wsrl{\eqalign{& G_{ij} = g_{ij} - (2{\pi}{\a}^{\prime})^2
\left( B g^{-1} B \right)_{ij} \cr & {\t}^{ij} =
-(2{\pi}{\a}^{\prime})^2 \left( {1\over{g + 2{\pi}{\a}^{\prime}
B}} B {1\over{g - 2{\pi}{\a}^{\prime}B}} \right)^{ij} \cr & g_{\rm
YM}^2 = 2{\pi} g_{s} \left( {\rm det} \left( 1 +
2{\pi}{\a}^{\prime} g^{-1}B \right)\right)^{\half}\ . \cr}}
Suppose the open string metric is Euclidean: $G_{ij} = {\d}_{ij}$,
then \bfi, \ \wsrl\ imply: \eqn\clstr{g = dx_3^2 +
{{(2{\pi}{\a}^{\prime})^2}\over{(2{\pi}{\a}^{\prime})^2 + {\t}^2}}
\left( dx_1^2 + dx_2^2 \right), \qquad B =
{{\t}\over{(2{\pi}{\a}^{\prime})^2  + {\t}^2}}\ , } and
\eqn\clscpl{g_{s} = g_{\rm YM}^2
{{{\a}^{\prime}}\over{\sqrt{(2{\pi} {\a}^{\prime})^2 + {\t}^2}}}\
. }

The Seiberg-Witten limit is achieved by taking ${\a}^{\prime} \to 0$ with
$G, {\t}, g_{\rm YM}^2$ kept fixed.

In this limit the effective action of the D3-brane theory will
become that of the (super)Yang-Mills theory on a noncommutative space
${\CA}_{\t}$. The relevant part of the
energy functional is:
\eqn\nwenrg{{\CE} = {1\over{2g_{\rm YM}^2}}
\int d^3 x \,
 {\Tr} \left( B_i \star B_i + {\nabla}_{i} {\Phi} \star
{\nabla}_i {\Phi} \right) \  ,}
where
\eqn\mgnf{B_i = {i\over 2} {\ve}_{ijk} F_{jk}\ .}
The fluctuations of the D3-brane in
some distinguished transverse direction (which we called
${\Phi}$) are described by the dynamics of the Higgs field.

As in the ordinary, commutative case, one can rewrite
\nwenrg\ as a sum of a total square and a total derivative:
\eqn\splt{{\CE} = {1\over{2g_{\rm YM}^2}}
\int d^3 x
 {\Tr} \left( {\nabla}_{i} {\Phi} \pm B_i \right)^2 \mp
{\p}_i {\Tr} \left( B_i \star {\Phi} + {\Phi} \star B_i  \right)}
The total derivative term depends only on the boundary conditions. So,
to minimize the energy given  boundary conditions
  we should  solve the {\it Bogomolny} equations:
\eqn\bgmlni{{\nabla}_i
{\Phi} \quad = \quad  \pm B_i, \quad i = 1,2,3\ .}

\subsec{Nahm's construction for commutative monopoles}
\noindent

To begin with, we review the techniques which  have been used in
the commutative case. Specifically,   Bogomolny equations take the form:
\eqn\bgmln{{\nabla}_i
{\Phi}   + B_i = 0, \quad i = 1,2,3}
{}The boundary condition is that
  at the  spatial infinity the Higgs field
approaches a constant, corresponding to the Higgs vacuum.
  In the case of $SU(2)$ this means   that locally on the
two-sphere at infinity: \eqn\asmt{{\phi} (x) \sim {\rm diag}
\left( {a\over 2}, - {a\over 2} \right)\ .} The solutions are
classified by the magnetic charge $k$. By virtue of the equation
\bgmln\ the monopole charge can be expressed as the winding number
which counts  how many times the two-sphere ${\bf S}^2_{\infty}$ at
infinity is mapped to the coset space $SU(2)/U(1) \approx {\bf
S}^2$ of the abelian subgroups of the gauge group.

Nahm \nahm\ constructs  solutions to
the monopole equations as follows. Consider the matrix
differential operator on the interval $I$
  with the coordinate
$z$:
\eqn\tmatr{-i{\Delta} =
  {\p}_{z} +  {\CT}_i {\s}_i, }
where
\eqn\tmatri{{\CT}_i = T_i(z) + x_i \ .}
$x_i$ are the coordinates in the physical space ${\bf R}^3$,
and the $k \times k$
matrices $T_i(z) = T_i^{\dagger} (z)$ obey Nahm's equations:
\eqn\nms{{\p}_{z} T_i = i{\ve}_{ijk} T_j T_k\ , }
with certain boundary conditions.
We take $I = ( -a/2, a/2)$ where $a$ is given in  \asmt.
At $z \to z_{0}$, $z_{0} = {\pm} a/2$ we require that :
\eqn\asmtt{T_i \sim {t_i \over z - z_{0}} + {\rm reg.}, \quad
[ t_i, t_j] = i {\ve}_{ijk} t_k\ ,}
i.e. the residues $t_i$ must form a $k$-dimensional
representation of $SU(2)$ (irreducible if the solution is to be non-singular).

\noindent
Then one looks for the fundamental solution to the equation:
\eqn\drc{-i {\Delta}^{\dagger} {\Psi} (z) =  {\p}_z {\Psi}
-  {\CT}_i {\s}_i {\Psi} = 0\ ,}
where
$${\Psi} =  \pmatrix{{\Psi}_{+} \cr {\Psi}_{-}},$$
and ${\Psi}_{\pm}$ are $k \times 2$ matrices ($k$ is the monopole
charge, and $2$ is for $SU(2)$), which must be finite at
$z = {\pm} a/2$ and normalized so that:
\eqn\nrmlz{\int dz \,\,  {\Psi}^{\dagger} {\Psi} = {\bf 1}_{2 \times 2}\ .}
(this $2 \times 2$ is again for $SU(2)$.)

\ndt
Then:
\eqn\sltn{\eqalign{A_i & \quad = \quad \int dz \, {\Psi}^{\dagger}
{\p}_i {\Psi}, \cr
{\Phi} & \quad = \quad  \int dz \, z \, {\Psi}^{\dagger} {\Psi}\ . \cr}}

\noindent
Notice that the interval $I$ could be $(a_1, a_2)$ instead of
$(-a/2, +a/2)$. The only formula that is not invariant under
shifts of $z$ is the expression \sltn\ for $\phi$.
By shifting $\phi$ by a scalar $(a_1 + a_2)/2$ we can make it
  traceless  and map $I$ back to the form we used above.

\subsec{Nahm's equations from the D-string point of view}

The meaning of the Nahm's equations becomes clearer in the D-brane
realization of   gauge theory and  the D-string construction of monopoles.
The endpoint of a fundamental string touching a D3-brane looks like
an electric charge for the $U(1)$ gauge field on the brane.
By S-duality, a D-string touching a D3-brane creates a magnetic
monopole. If one starts with two parallel D3-branes, seperated by
  distance $a$ between them, one  is studying the $U(2)$
gauge theory, Higgsed down to $U(1) \times U(1)$,
where the vev of the Higgs field is
$$
{\Phi} = \pmatrix{ a_1 & 0 \cr 0 & a_2 \cr}
$$
One can suspend a D-string between these two D3-branes, or a collection of
$k$ parallel D-strings. These would correspond
to a charge $k$ magnetic monopole in the Higgsed $U(2)$ theory.
The BPS configurations of these D-strings are described the
corresponding self-duality equations in the 1+1 dimensional $U(k)$
gauge theory on the worldsheet of these strings \manuel. The equations
\nms\ are exactly these BPS equations. The presence of the D3-branes
is reflected in the boundary conditions \asmtt.
The matrices $T_i$ correspond to the ``matrix'' transverse coordinates
$X^i$, $i=1,2,3$ to the D-strings, which lie within D3-branes.
\subsec{Charge one monopoles}

In the case $k=1$ the analysis simplifies:
$T_i = 0$, and
\eqn\fndm{{\Psi} = \pmatrix{ \left( {\p}_z + x_3 \right) v \cr
\left( x_1 + i x_2 \right) v }, \quad
{\p}_z^2 v = r^2 v, \quad r^2 = \sum_i x_i^2 \ .}
The condition that $\Psi$ is finite at both ends of the interval
allows two solutions for \fndm:
$$
v = e^{\pm r z},
$$
which after imposing the normalization condition,\nrmlz, leads to:
$$
{\Psi} = {1\over \sqrt{2{\rm sinh} (ra)}} \pmatrix{
\sqrt{r + x_3} e^{rz} & - \sqrt{r - x_3} e^{-rz} \cr
{x_{+} \over \sqrt{r + x_3}} e^{rz} & {x_{-} \over \sqrt{r - x_3}} e^{-rz} },
$$
where we used $x_{\pm} = x_1 {\pm} i x_2$.

In particular,
$$
{\Phi} = {1\over 2}\left( {a\over  {\rm tanh} (ra)} - {1\over r}
\right) {\s}_{3}.
$$

\subsec{Abelian ordinary monopoles} \ndt It is interesting that
Nahm's equations describe Dirac monopoles as well. To achieve this
  replace the
interval $(-a/2, a/2)$ by the interval $( - \infty, a)$.
Intuitively this is natural, since in the $U(1)$ case the Higgs
field $\phi$ has only one eigenvalue at infinity.

Then the equation \bgmln\ becomes simply the condition that the
abelian monopole has a magnetic potential $\phi$, which must be
harmonic. Let us find this harmonic function. The matrices $T_i$
can be taken to have  the following form: \eqn\sutwo{T_i (z) = {t_i
\over z }, \quad [t_i , t_j] = i {\ve}_{ijk} t_k \ ,} where $t_i$
form an irreducible spin $j$ representation of $SU(2)$. Let $V
\approx {\IC}^{N}, \,  N = 2j+1$, be the space of this representation.
The matrices $\Psi_{\pm}$ are now $V$-valued. By an $SU(2)$
rotation we can bring the three-vector $x_i$ to the form
$(0,0,r)$, i.e. $x_1 = x_2 = 0, x_3 > 0$. Then one can show that
in this basis \eqn\ps{{\Psi}_{-} = 0, \quad {\Psi}_{+} =  {\n}_{j}
z^{j} e^{rz} \vert j \rangle \ ,} where $\vert j \rangle \in V$ is the
highest spin state in $V$. The coefficient ${\n}_j$ is found from the
normalization condition: \eqn\cj{\vert {\n}_j \vert^2 = {{r^{2j+1}
}\over {(2j)!}}\ .}

\noindent From this we get the familiar formula for the singular
Higgs field \eqn\sing{ \phi = a - {N\over 2r}\ .} \noindent

\newsec{Abelian noncommutative monopoles}

In this section we study the solutions to the Bogomolny
equations for  U(1) gauge theory on a noncommutative three
dimensional space. As before, we
assume the Poisson structure ($\t$) which deforms the multiplication of
the functions to be constant. Then there is essentially a unique
choice of  coordinate functions $x_1, x_2, x_3$ such that the
commutation relations between them are as follows:
\eqn\ncm{\eqalign{[ x_1, x_2 ] = & - i{\t}, \quad {\t}
> 0\cr
[x_1 , x_3 ] = & [x_2, x_3 ] = 0 \ . \cr}}This  algebra, \ncm, defines
noncommutative ${\bf R}^3$, which  we denote   by
${\CA}_{\t}$. Introduce the creation and annihilation operators $c,
c^{\dagger}$: \eqn\osci{c = {1\over{\sqrt{2\t}}} \left( x_1 - i
x_2 \right), \quad c^{\dagger} = {1\over{\sqrt{2\t}}} \left( x_1 +
i x_2 \right)\ , } that  obey $$ [c , \, c^{\dagger} ] = 1.$$

\subsec{Noncommutative Nahm equations}

\noindent We start by repeating the  procedure \neksch\ that worked in the
ADHM instanton case, namely we relax the condition that $x_i$'s commute but
insist on the equation \nms\ with $T_i$ replaced by the relevant
matrices ${\CT}_i = T_i + x_i$. Then the equation \nms\ on $T_i$
is modified: \eqn\mdnms{{\p}_z T_i = i {\ve}_{ijk} T_j T_k +
{\d}_{i3} {\t}\ .} It is obvious that, given a solution $T_i(z)$ of
the original Nahm equations, it is easy to produce a solution of
the noncommutative ones: \eqn\mdf{T_{i}(z)^{\rm nc} = T_{i}(z) +
{\t} z {\d}_{i3}\ . } From this it follows that, unlike the case  of instanton
moduli space, the monopole moduli space {\it does not change under
noncommutative deformation}.

This deformation, \bak\mdnms\  , is   exactly what one gets by looking
at the D-strings suspended between the D3-branes (or a semi-infinite
D-string with one end on a D3-brane) in the presence of a $B$-field.
One gets the deformation:
\eqn\dfrm{
[X^i, X^j] \to [X^i, X^j] - i{\t}^{ij}
= [T_i, T_j] - {1\over 2} {\t} {\ve}_{ij3}}
The reason why  ${\t}^{ij}$, instead of $B_{ij}$, appears
on the right hand side of \dfrm\ is rather simple. By applying T-duality in
the directions
$x_1, x_2, x_3$ we could map
the D-string into the D4-brane. The matrices $X^1, X^2, X^3$ become
the components $A_{\hat 1}, A_{\hat 2}, A_{\hat 3}$ of the gauge field on
the D4-brane
worldvolume, and the $B$-field would couple to these
gauge fields via the standard coupling
$F_{{\hat i}{\hat j}} - {\hat B}_{{\hat i}{\hat j}}$, where
${\hat B}_{{\hat i}{\hat j}}$
 is the T-dualized $B$-field. It remains to observe that
${\hat B}_{\hat i \hat j} = {\t}^{ij}$, since the T-dualized indices
${\hat i}$ label the coordinates on the space, dual to that of
$x_i$'s.

\subsec{Solving Nahm's equations}

\ndt{}To
solve \mdnms\  we imitate the approach for the $k=1$ commutative monopole
by taking \eqn\strng{T_{1,2}
= 0, \, T_{3} = {\t} z.}
To solve \drc\  for  $\Psi$ we introduce the operators $b, b^{\dagger}$:
\eqn\cran{ b = {1\over{\sqrt{2\t}}} \left( {\p}_z +  x_3 + {\t} z
  \right), \quad \quad b^{\dagger} = {1\over{\sqrt{2\t}}} \left( -
{\p}_z +  x_3 + {\t} z  \right)\ , } which obey the oscillator
commutation relations: \eqn\osc{ [ b, b^{\dagger} ] = [ c,
c^{\dagger} ] = 1\ .}  We introduce  the {\it superpotential} \eqn\sprn{ W
= x_3 z + {1\over 2}{\t} z^2\ , } so that  $b = {1\over{\sqrt{2\t}}}
e^{-W} {\p}_z e^{W}, \quad b^{\dagger} = - {1\over{\sqrt{2\t}}}
e^{W} {\p}_z e^{-W}$. Then equation \drc\ becomes:
\eqn\drcnc{\eqalign{b^{\dagger}  \Psi_{+} +  &  c {\Psi}_{-} =
0\cr c^{\dagger} {\Psi}_{+} - & b {\Psi}_{-} = 0 \ .\cr}}

\ndt  The general solution of \drc\  is:
  \eqn\drcii{{\Psi}_{+} = \left( u_1 b + u_2 c \right) v,
\quad {\Psi}_{-} = \left( u_1 c^{\dagger} - u_2 b^{\dagger}
\right) v\ ,} where $v$ satisfies
\eqn\mstrr{ (b^{\dagger}b+cc^{\dagger})v=0 \ ,}
and $(u_1, u_2)$ are two
complex numbers defined up to   multiplication by a common factor
(which can be reabsorbed in the definition of $v$).
Thus, the generic solution is parameterized by a point $u = (u_1 : u_2) \in
{\bf CP}^1$ on a two-dimensional (twistor)  sphere.

To construct the solution we must now solve \mstrr\ and normalize
$\Psi^\dagger \Psi$.
Recall that $z$ is defined    on the half line
$(-\infty, a)$,  so therefore $b$ and $b^{\dagger}$ are not Hermitian
conjugates: \eqn\hrmt{\int {\psi}^{*}_{1} b \psi_{2} =
{\psi}_{1}^{*} {\psi}_{2} (z=0) + \int \left( b^{\dagger} \psi_{1}
\right)^{*} {\psi}_{2}\ .} As far as the $c, c^{\dagger}$ system is
concerned we  will  work in the occupation number basis of
the  Fock space obtained by
quantizing the $(x_1, x_2)$ plane.
The most general expression for $v$  is:
\eqn\gnans{v = \sum_{n=0}^{\infty} v_{n m} (z, x_3) \vert n \rangle \langle
m \vert \ .}
However it  turns out that when  $u = (1:0)$ or when $u = (0:1)$ one can
make the following ansatz (which is equivalent to imposing   axial symmetry
on $v$):
\eqn\axlans{ v = \sum_{n = 0}^{\infty} v_{n} (z, x_3) \vert n
\rangle \langle n \vert, \quad cc^{\dagger} \, \vert n\rangle =
(n+1) \, \vert n \rangle \ .} In this case  \mstrr\  becomes
\eqn\oscil{ b^{\dagger} b v_{n}(z , x_3) = - (n+1) v_{n} (z, x_3)\ , }
in the class of functions that  lead to a $\Psi$  that is
normalizable on  the half  line $(-\infty, a)$.
It is obvious that if we can solve for $v_0$ then $v_n  \propto b^nv_0$;
so the solution of \mstrr\ is
\eqn\nunol{v_{n}   =   {\nu}_{n} b^{n} {\vf}(z)\ ,}
where
\eqn\phisol{{\vf}(z)
\equiv   e^{W(z)} \int_{- \infty}^{z} e^{-2W (p)} dp \, }
and
${\nu}_{n}$ are normalization constants to be determined below.
Will this lead to a normalizable $\Psi$?
As $z \to - \infty$ we have the  estimate ${\vf} (z) < e^{-W(z)}$,
so  that $v_0$, as well as all its descendants $v_{n}$,  are good functions.
Notice that we can only make it normalizable  on a half line,
which nicely fits with the intuition that the abelian Higgs field
must have  values that are bounded.
\medskip
\noindent $\bullet$ Notice that by shifting the coordinate $x_3
\mapsto x_3 + {\t} a$ we can always make $a = 0$ (this is
impossible for ${\t} = 0$). From now on we assume $a = 0$.

\noindent $\bullet$ After this shift we see that the only
dimensional parameter in the problem is ${\t}$. Let us choose
the length units in which $2{\t}=1$.

\noindent In the case, where   $u$ is a   generic
point on the
two sphere, we have
$$
v = \sum_{n,m} {\n}_{nm} b^{n} {\vf} (z) \vert n\rangle \langle m \vert \ .
$$
In this paper we shall only discuss the case where  either $u = (1:0)$,
or $u = (0:1)$.

\subsec{The normalization condition}

We   start by considering  the case $u = (0:1)$.
Accordingly,   ${\Psi}_{+} = c v, \quad {\Psi}_{-} = - b^{\dagger}v$, and
$$ {\Psi}^{\dagger} {\Psi} = \sum_{n=0}^{\infty} \left[ \left(
b^{\dagger}  v_n \right)^{\dagger} \left( b^{\dagger}v_{n} \right) +
n \vert v_{n} \vert^2
\right]
\vert n \rangle \langle n \vert = - \sum_{n=0}^{\infty} {\p}_{z}
\left( v_{n}^{\dagger} b^{\dagger} v_{n} \right) \vert n \rangle \langle n
\vert \ .  $$ The noncommutative version of the condition
\nrmlz\ is:
\eqn\nrmlznc{\int_{-\infty}^{0} dz\,\, {\Psi}^{\dagger} {\Psi}
= 1 = \sum_{n=0}^{\infty} \vert n \rangle \langle n \vert \ , }
thus
\eqn\norm{ \left( v_{n}^{\dagger} b^{\dagger} v_{n} \right)(z=0) =-1 \ .}
 which
reduces to the sequence of relations: \eqn\nr{
n \vert\nu_n\vert^2 ({\p}^{n} {{\u}(2x_3)}) ({\p}^{n-1}
  {\u}(2x_3)) = 1}
where
\eqn\mstr{\eqalign{
{\u}(z) \quad = & \quad
  \int_{0}^{\infty} e^{-{p^2 \over 2} + z p} dp \cr
& = \sqrt{\pi \over 2} e^{z^2 \over 2} \left( 1+
\ee{{z\over\sqrt{2}}} \right) \cr & = \sum_{n=0}^{\infty}
{{\left( {{n-1}\over{2}} \right) !}\over{n!}} 2^{{n-1}\over 2} z^n
\cr & \sim \sqrt{2\pi} e^{z^2 \over 2}, \quad z \to + \infty\cr &
\sim {1\over{\vert z \vert}}, \quad \quad z \to - \infty \ .\cr}}
For $n=0$ \nr\  is  explicitly given by the analytic continuation of \nr\
to $n=0$,
namely $\vert\nu_0\vert^2 {\u}(2x_3)=1 \ . $
The
function \mstr\ obeys the following differential equation
\eqn\dfe{{\p}_z {\u}(z)  = z {\u}(z) + 1, \qquad {\u}(0) =
\sqrt{{\pi}\over 2}\ .} Introduce the expansion coefficients
\eqn\zfun{\eqalign{{\u} (2x_3 + y) \quad = \quad & \sum_{n=0}
{\z}_{n} {{y^{n}}\over{n!}}\ ,\cr {\z}_{n} \quad = \quad &
\int_{0}^{\infty} p^{n} e^{-{p^2 \over 2} + 2px_3} dp\ ,\cr}} which
obey the following equations:
\eqn\zfunn{\eqalign{
  {\z}_{n+1}
\quad = \quad & 2x_3 {\z}_n + n {\z}_{n-1}\cr {\p}_{3} {\z}_n
\quad = \quad & 2{\z}_{n+1} \cr {\z}_n ( x_3 = 0) \quad = \quad &
2^{{n-1}\over 2} \left( {{n - 1}\over 2} \right)!\ \ . \cr
  }}
The recursion relation  in \zfunn\ for $n=0$ is to be understood
by analytic continuation  as $n \to 0$. In this limit we have  $n{\z}_{n-1}
\to 1$, as $n \to 0$.
Thus ${\z}_1 =2x_3{\z}_0+1$, as can also be checked directly from \zfun.

To find the normalization constants we  substitute
\zfun\ into \nr\ to deduce:
\eqn\nrmlz{ \vert {\nu}_{n}
\vert^2 = {1\over{n {\z}_{n}  {\z}_{n-1}}}\ . }
Again, for $n=0$
the last equality
is understood as $\vert {\nu}_{0}
\vert^2= 1/\z_0$.
This completes the solution. We have explicitly constructed $v$
and thus $\Psi_\pm$, from which we can determine, using  \sltn,
the Higgs and gauge fields. To do this we shall
need to evaluate the
overlap integrals:  \eqn\ovrlp{\int^{0}_{-\infty} (b^{n} {\vf}
)(b^{n+1} {\vf}) = {\z}_{n+2}{\z}_{n} - {\z}_{n+1}^2 =
(n+1){\z}_n^2 - n {\z}_{n+1}{\z}_{n-1}\ . }Again, for $n=0$
the last equality
is understood with $n {\z}_{n-1} = 1$ for $n=0$.

Let us also introduce the functions ${\xi}, {\tilde \xi}\ $ and
${\eta} = {\tilde\xi}^2$ :
\eqn\etaxi{{\tilde\xi} ( n ) = \sqrt{{\z}_{n} \over {\z}_{n+1}},
\qquad
{\eta} (n) =
{{\z}_{n} \over{{\z}_{n+1}}}, \qquad {\xi} (n) = \sqrt{n{\z}_{n-1} \over
{\z}_{n}}
\ . }
We will need the asymptotics of these functions for large $x_3$.
  Let $r^2_{n} = x_3^2 + n$. For  $r_{n} + x_3 \to \infty$  we
can estimate the integral in \zfun\ by the saddle point method.
The saddle point and the approximate values of ${\z}_n$ and $\eta_n$  are:
\eqn\sdlpnt{\eqalign{{\bar p} \quad = \quad &  x_3 + r_{n}\cr {\z}_{n} \quad
\sim \quad \sqrt{{\pi}\over{r_{n}}} & \left( x_3 + r_{n} \right)^{n +
{1\over 2}} e^{{1\over{2}} \left(  x_3 + r_{n} \right)\left( 3x_3 - r_{n}
\right)}\cr {\eta}_{n} \sim \quad {1\over{x_3 + r_{n+1}}} & \left( 1 + {1\over
{4r^2_{n}}} + \ldots \right)\ .  \cr}}

\medskip

\subsec{The explicit solution for the gauge and Higgs fields}
\sssec{\rm The \quad Higgs \quad Field.}

\ndt{}The Higgs field is given by  \sltn: \eqn\hgs{{\Phi} =
\int dz \, z \, \Psi^\dagger \Psi \equiv\sum_{n=0}^{\infty} {\Phi}_{n}
(x_3) \vert n \rangle \langle n \vert \ , } it has   axial
symmetry, that is commutes with the
the number
operator $c^{\dagger}c$. Explicitly:
\eqn\hgsn{\eqalign{{\Phi}_{n}
\quad = \quad &  \int v_{n}^{*}b^{\dagger} v_{n} dz = \cr & =
{{{\z}_{n}}\over{{\z}_{n-1}}} - {{{\z}_{n+1}}\over{{\z}_{n}}} =
\, {\p}_{3} {\rm log} {\xi}_{n} \cr & = (n-1) {\eta}_{n-2} - n
{\eta}_{n-1}, \qquad n > 0 \cr & = - {{\z}_1 \over {\z}_0} = -2x_3-{1 \over
{\z}_0} ,  \qquad\quad n =0\ . \cr}}
{}To arrive at the third line we used the
fact that
$$
{1\over \eta_n}-{1\over \eta_{n+1}} = n \eta_{n-1} -(n+1) \eta_{n} \ ,
$$
which follows immediately from the recursion relation for the $\z'$s in
\zfunn. {}These fields are finite at $x_3=0$. Indeed as $x_3
\to
0$,
\eqn\hgszr{
{\Phi}_{n} (x_3 = 0)\quad  =
\sqrt{2} \left( {{\left( {{n-1}\over 2}
\right)!}\over{\left({{n-2}\over 2} \right)!}} - {{\left(
{{n}\over 2} \right)!}\over{\left({{n-1}\over 2} \right)!}} \right) \ .}
At the origin:
\eqn\orign{{\Phi}_{0} (x_3 = 0)\quad  = \quad  - \sqrt{2\over \pi}\  .}
\sssec{\rm The \quad Gauge \quad Field.}

\ndt{}Using \sltn\ it is easy to see that the component $A_3$ vanishes
\eqn\athree{\eqalign{A_3
\quad =  \quad & \int ( b^{\dagger}v_n)^{\dagger}
{\p}_3 (b^{\dagger}v_n) + n v_{n}^{\dagger}
{\p}_3 v_n = \cr & -
(b^{\dagger}v_n)^{\dagger} {\p}_3 v_n (z=0) + \int (b^{\dagger} v_n)^{\dagger}
v_n = \cr & {\p}_3 {\rm log} {\xi}({n})  + {{{\z}_{n+1}{\z}_{n-1} -
{\z}_{n}^2}\over{{\z}_{n-1} {\z}_{n}}} = 0 \ .
  \cr}}

\ndt{}
In the same gauge the components $A_1, A_2$ (which we consider to
be anti-hermitian) are given by:
\eqn\ggef{\eqalign{A_{c} =
{1\over 2} \left( A_1 + i A_2 \right), \quad & A_{c^{\dagger}} =
{1\over 2} \left( A_1 - i A_2 \right) = - A_{c}^{\dagger} \cr
A_{c} \quad = \quad  & \int {\Psi}^{\dagger} [ {\Psi}, c^{\dagger}
]  \cr =  & \quad {\xi}^{-1} [ {\xi}, c^{\dagger} ] = c^{\dagger}
\left( 1 - {{\xi}(n) \over {\xi}(n+1)} \right) \ . \cr}
}
Again we see that the matrix elements of $A_c$ are all finite and non singular.

\sssec{\rm The \quad   Field \quad strength.}

\ndt{}
From \ggef\ we deduce:
$$
F_{12} = 2i\,
 \left( {\p}_{c}A_{c^{\dagger}} - {\p}_{c^{\dagger}} A_{c} + [ A_{c},
A_{c^{\dagger}} ] \right) =
$$
\eqn\fonetwo{\eqalign{
& 2 \,
\left( \left[ {{{\xi} (n)}\over{{\xi}(n+1)}} c,
c^{\dagger} {{{\xi} (n)}\over{{\xi}(n+1)}}  \right] - 1 \right) = \cr
& = 2 \,
\sum_{n > 0}
\left( - 1 + (n+1) \left( {{{\xi}(n)}\over{{\xi}(n+1)}} \right)^2 -
n \left( {{{\xi} (n-1)}\over{{\xi}(n)}} \right)^2 \right) \vert n \rangle
\langle n \vert + \cr
& \qquad\qquad\qquad +   2\,  \left( - 1 +
\left( {{{\xi}(0)}\over{{\xi}(1)}} \right)^2 \right) \vert 0
\rangle \langle 0 \vert  \  , \cr}}
from which it follows, that: \eqn\flds{\eqalign{B_3 (n)
\quad = \quad
& 2 \left(1 - n {{{\eta}_{n-1}}\over{{\eta}_{n}}}
   + \left(n-1\right)
{{{\eta}_{n-2}}\over{{\eta}_{n-1}}}
  \right)\cr
B_{c} \quad = \quad & {1\over 2} \left( B_1 + i B_2 \right) =
c^{\dagger} {{{\xi} (n) }\over{{\xi} (n+1) }} \left( {\Phi} (n) -
{\Phi} (n+1) \right)\ .\cr}}
with the understanding that at $n=0$:
\eqn\batz{  B_3(0) = 2\left(1 - {\z_1\over \z_0^2}\right) .   }
\subsec{ Checking  the  Bogomolny equations.}

\ndt{}
With
our conventions it is relatively easy to check that our solution satisfies
the Bogomolny equations everywhere:
\eqn\chck{\eqalign{& {\nabla}_{3} {\Phi} = {\p}_3 {\Phi} = - B_3 \cr &
{\nabla}_{c} {\Phi} = {\xi}^{-1} {\p}_{c} {\Phi} {\xi} = -B_{c} \ . \cr}}
For example, consider the equation ${\p}_3 {\Phi} = - B_3 $.
We have:
$$ {\p}_3 {\Phi}(n)= {\p}_3 [(n-1)\eta_{n-2}-n\eta_{n-1}].$$
Then we use the fact that
\eqn\deret{\p_3\eta_n=2\left(1-{\eta_{n}\over \eta_{n+1}} \right)\  , }
to see that
\eqn\check{\eqalign{&{\p}_3 {\Phi}(n) = 2(n-1)\left(1-{\eta_{n-2}\over
\eta_{n-1}} \right)-
2n\left(1-{\eta_{n-1}\over \eta_{n}} \right) = \cr
&  -2 \left(1-n{\eta_{n-1}\over
\eta_{n}}+(n-1)
{\eta_{n-2}\over \eta_{n-1}} \right) = -B_3(n) . \cr}}

\ndt\sssec{\rm Other  \quad solutions \quad and \quad Seiberg-Witten
\quad map.}

\ndt
It is plausible that the solutions corresponding
to the other values of $u = (u_1: u_2)$ also have a physical meaning.
In fact, the solutions of the Dirac-Born-Infeld theory
\moriyama\mateos\genmnp\ that
describe a  a D-string touching a D3-brane (or D-string suspended between
two D3-branes) in the presence of the $B$-field
suggest that ${\Phi}$ is multi-valued.
Moreover, the solution for $\Phi$ \moriyama\ is implicit,
whereas our solution is explicit. On the other hand the Seiberg-Witten map
from the noncommutative gauge fields to the commutative ones \witsei\ must
map our explicit solution for $(A, {\Phi})$ into
the solution of the DBI theory.
It could
mean that our solution is just one branch of the full solution,
somehow incorporating other choices of $u$. However,
we have found
that the solution corresponding to the choice $u = (1:0)$
does not quite satisfy the BPS equations everywhere. Instead, it has a
source, localized
along a  semi-infinite string pointing in the $x_3 \to -\infty$ direction.
Nevertheless, it is clear that the other
``branches'', corresponding to the generic $u$,  seem worth investigating
further.  It is also plausible that in order to have a better
understanding of the matching of the solutions to the DBI theory and the
noncommutative gauge theory one would need to incorporate the
${\a}^{\prime}$
corrections.

\ndt\sssec{\rm  A\quad  remark \quad concerning \quad instantons.}

\ndt
As in the
ordinary gauge theory case the monopoles  are the solutions of  the
instanton equations in four dimensions, that are invariant under
translations in the fourth direction $x_4$.  We observe that the
solution  presented above (\hgsn,\ggef\ ), can also be cast in the Yang
form: Take $\xi = \xi (x_3, n)$ as in \ggef. Then ${\p}_3 {\xi}$
commutes with ${\xi}$ and we can write ${\p}_3 {\xi} {\xi}^{-1} =
{\p}_{3} {\rm log} {\xi}$. The formulae \ynga\ yield exactly
\ggef\ and \hgsn\ with ${\Phi} = i A_{4}$. Indeed, the equation
\ynge\ is nothing but the first equation in \chck.

\subsec{Toda lattice}

At this point it is worth mentioning the relation of the
noncommutative Bogomolny equations with the Polyakov's non-abelian
Toda system (see \mtoda\ for the recent studies of this system) on
the semi-infinite one-dimensional lattice. Let us try to solve the
equations \bgmln\ using the Yang ansatz and imposing the axial
symmetry: we assume that ${\xi} ( x_1, x_2, x_3 ) = {\xi} (n ,
x_3)$, $n = c^{\dagger}c$. Then the equation \ynge\ for the
$x_4$-independent fields reduces to the system: \eqn\toda{
{\p}_{t} ( {\p}_{t} g_n g_n\inv )  - g_{n}g_{n+\scriptscriptstyle
1}\inv + g_{n-\scriptscriptstyle 1} g_{n}\inv = 0} where $$
g_{n}(t) = {{e^{{t^2 \over 2}}}\over{n!}} {\xi}^2 \left(n, {t\over
2}\right)\ , $$ (notice that $g_{n}(t)$ are ordinary matrices). In
the $U(1)$ case we can write $$ g_{n}(t) = e^{{\a}_{n}(t)}\ , $$
and rewrite \toda\ in a more familiar form: \eqn\todai{{\p}_{t}^2
{\a}_{n} + e^{{\a}_{n-1} - {\a}_{n}} - e^{{\a}_{n} - {\a}_{n+1}} =
0}For $n = 0$ these equations also formally hold if we set $g_{-1}
= 0$ (this boundary condition follows both from the Bogomolny
equations and the same condition is imposed on the Toda variables
on the lattice with the end-points).

Our Higgs field ${\Phi}_{n}$ has a simple relation to the
${\a}$'s:
$$
{\Phi}(x_3, n) = - 2x_3 + {\a}_{n}^{\prime} (2x_3) \ .
$$
Our solution to \todai\ is:
\eqn\ours{{\a}_{n} = {1\over 2} t^2 + {\rm log}
\left( {{n{\z}_{n-1}(t/2)}\over{{\z}_{n}(t/2)}} \right) - {\rm log} ( n! )\ .}

It is amusing that Polyakov's motivation for studying   the
system \toda\ was  the structure of
loop equations for   lattice gauge theory. Here we encountered these
equations in the study of the {\it continuous}, but
noncommutative, gauge theory, thus giving more evidence for their
similarity.

We should note in passing that in the integrable non-abelian Toda system
one usually has two `times' $t, {\bar t}$,
so that the equation \toda\ has actually the form \mtoda:
\eqn\todaii{ {\p}_{t} ( {\p}_{\bar t} g_n g_n\inv )  -
g_{n}g_{n+\scriptscriptstyle 1}\inv +  g_{n-\scriptscriptstyle 1} g_{n}\inv
= 0\ .}

It is obvious that these equations describe four-dimensional axial
symmetric instantons
on the noncommutative space with the
 coordinates $t, {\tb}, c, c^{\dagger}$ of which only half
is noncommuting.

\subsec{The mass of the monopole}

\ndt{} In this section we restore our original units, so that
$2{\t}$ has dimensions of (length)$^2$. From the formulae \sdlpnt\
we can derive the following estimates: \eqn\asmphi{ {\Phi} (n)
\sim -{1\over{2r_{n}}}= -{1\over{2\sqrt{x_3^2+2 {\t} n}}}\quad n
\neq 0, \, r \to \infty \ .} Instead, for $n=0$ we have:
\eqn\asm{\eqalign{& {\Phi} (0) \sim - {{x_3}\over{\t}}, \qquad x_3
\to + \infty \cr &{\Phi} (0) \sim -{1\over{2\vert x_3 \vert}},
\qquad x_3 \to - \infty \ .\cr}} {}The  asymptotics of the
magnetic field is clear from the Bogomolny equations and the
behaviour of $\Phi$. Thus, for example, \eqn\asyb{B_3(n) =
-\partial_3 \Phi(n) = -{x_3\over 2r_{n}^3} ,\quad n\neq0  , } and
similarly for the other components of $B$. This is easily
translated into ordinary position space, as in the discussion
following equation\mps, since, for large $n$,    $B_i(n, x_3) \sim
B_i( x_1^2+x_2^2 \sim n, x_3)$. Therefore the magnetic field for
large values of $x_3 $ and $n$, or  equivalently large $x_i$ is
that of a point-like magnetic charge at the origin. However the
$n=0$ component of $B_3$ behaves differently for large positive
$x_3$: \eqn\Bthras{ B_3(n=0) = -\p_3\Phi(0) = {1\over{\t}} \ .}
Notice, that this is exactly the value of the $B$-field. Thus, in
addition to the magnetic charge at the origin we have a flux tube,
 localized in a Gaussian packet in the $(x_1 ,x_2)$ plane, of the
size $\propto \theta$, along the positive
$x_3$ axis. The monopole solution is indeed a smeared version of the Dirac
monopole,
wherein the Dirac string (the D-string!) is physical.

To calculate the energy of the monopole we use the  Bogomolny
equations to reduce the total energy to a boundary term:
\eqn\msm{\eqalign{{\CE} = &
{1\over{2g_{\rm YM}^2}} \, \int d^3 x \,
\left( {\vec B} \star {\vec B} + {\vec\nabla} {\Phi} \star {\vec\nabla}
{\Phi} \right)
  = \cr
& {1\over{2g_{\rm YM}^2}} \, \int d^3 x \left( {\vec B} + {\vec\nabla} \Phi
\right)^2  - {1\over{2g_{\rm YM}^2}}
\, \int d^3 x \, {\vec\nabla} \cdot
\left( {\vec B} \star {\Phi} + {\Phi} \star {\vec B}
\right) = \cr
&  {{2\pi {\t}}\over{2g_{\rm YM}^2}}
\int dx_3 \sum_{n} \langle n \vert  {\p}_3^2 {\Phi}^2 +
4 {\p}_{c} \left( {\xi}^2 \left( {\p}_{c^{\dagger}} {\Phi}^2 \right)
{\xi}^{-2}\right) \vert n \rangle \ ,
\cr}}
where in the last line we switched back to the Fock space, by using the
relation:
\eqn\fcksp{\int dx_1 dx_2  f(x_1, x_2 )  = 2{\pi}{\t} {\Tr}_{\CH} {\hat f}}
Thus, the energy is given by the   boundary term. To evaluate this expression
we need to figure out what the boundary term is in the noncommutative,
Fock space setup?

Consider   the derivative  terms in \msm\ involving $x_1, x_2$.
They can be expressed as the commutators with $c$ or  $c^{\dagger}$.
In computing the trace
\eqn\trc{
{\Tr}_{\CH} [ c, {\CX} ]
= \sum_{n} \langle n \vert [ c, {\CX} ] \vert n \rangle \ ,}
where we denote by ${\CX}$ the terms ${\Phi} B_{c} + B_{c} {\Phi}$
in \msm\ ,  we get naively  get that the trace of a commutator is zero.
But we should be careful, since the  matrices are infinite and the
trace is an  infinite sum.
If we regulate it by restricting  the sum to $n \leq N$, then
the matrix element $\langle N  \vert c \vert N+1 \rangle
\langle N+1 \vert {\CX} \vert N \rangle$ is not cancelled, so that the
regularized trace is
\eqn\trcrg{{\Tr}_{{\CH}_{N}} [ c, {\CX} ]  = \sqrt{N+1} \langle N+1
\vert {\CX} \vert N \rangle}
and similarly for $c^{\dagger}$.
Let us  choose as  the infrared regulator box the ``region'' where
$\vert x_3 \vert \leq L, 0 \leq n \leq N$, $L \sim \sqrt{2{\t}N} \gg 1$.  Then
the total integral in \msm\ reduces to the sum of two terms
(up to the factor ${\pi {\t}}\over{g_{\rm YM}^{2}}$):
\eqn\msmi{\eqalign{&
4 N \int_{-L}^{L} dx_3 \,
{{{\eta}_{N-1}}\over{{\eta}_{N}}}
\left( {\Phi}_{N}^2 - {\Phi}_{N+1}^2 \right)\cr & \qquad \qquad
+\sum_{n=0}^{N} {\p}_3 {\Phi}_{n}^2
\vert^{x_3 =
+L}_{x_3 = - L} \ .\cr}}
The first line in \msmi\ is easy to evaluate. Since
$x_3 + r_{n} \geq -L + \sqrt{ L^2 + 2{\t} N } \propto \sqrt{{\t}N}
\to \infty$ we can use the asymptotic expressions \sdlpnt\ to
make an estimate:
$$
4N \int_{-L}^{L} dx_3 \,
{{{\zeta}_{N+1} {\zeta}_{N-1}}\over{{\zeta}_{N}^2}}
\left( {\Phi}_{N}^2 - {\Phi}_{N+1}^2 \right) \approx
\int_{-L}^{L} dx {{2{\t}N}\over{(x^2 + 2{\t}N)^2}}
$$
$$
\approx{{L}\over{L^2 + 2{\t} N}} \to 0
$$
The second line in \msmi\ contains derivatives
of the Higgs field evaluated at $x_3 = L \gg 0$ and
 at $x_3 = -L \ll 0$. The former is estimated using the
$z \gg 0$ asymptotics in \mstr\ or \sdlpnt, and produces:
$$
\sum_{n=0}^{N} {\p}_3 {\Phi}^{2}_{n} (x_3 = L) \sim {{2{\t}(N-1)}\over{L^3}} +
2{{L}\over{{\t}^2}}
$$
The diverging with $L$ piece comes solely from the $n=0$ term.
Finally, the $x_3 = -L$ case is treated via $z \ll 0$ asymptotics
in \mstr\ yielding the estimate $\sim {\t}N/L^{3}$ vanishing in the limit of
large $L, N$.

Hence the total energy is given by \eqn\ttlms{ {\CE} \propto
{{2\pi {\t} \times 2 L}\over{2g_{\rm YM}^2 {\t}^2}} = {2{\pi} L
\over{g_{\rm YM}^2 {\t}}} \ ,} which is the mass of a string of
length $L$ whose tension is $$ T = {{2{\pi}\over{g_{\rm YM}^2
{\t}}}} \ . $$

\subsec{Magnetic charge}

It is instructive to see what is the magnetic charge of our solution.
On the one hand, it is clearly zero:
\eqn\chrn{
Q \propto \int_{{\p} ({\rm space})} {\vec B} \cdot d{\vec S} =
\int d^3 x \,\,  {\vec\nabla} \cdot {\vec B} = 0
}
since the gauge field is everywhere non-singular. On the other hand,
we were performing a  $\t$-deformation of the Dirac monopole, which definitely
had magnetic charge.
To see what has happened let us look at \chrn\ more carefully. We  again
introduce the box and evaluate the boundary integral \chrn\ as in \trcrg:
\eqn\chnri{{Q\over{2\pi}} = \sum_{n=0}^{N} \left[ B_{3} (x_3 = L, n) -
B_{3} (x_3 = -L, n) \right]
+ 4N \int_{-L}^{L} dx_3 {{\eta}_{N-1} \over {\eta}_{N}}
\left( {\Phi}_{N} - {\Phi}_{N+1} \right)}
It is easy to compute the sums
\eqn\sms{\eqalign{& \qquad \qquad \qquad
\sum_{n=0}^{N} B_{3} (x, n) =  {\p}_3 {{\z}_{N+1} \over {\z}_{N}} \cr
& 4 N \int_{-L}^{L} dx_3 {{\eta}_{N-1} \over {\eta}_{N}}
\left( {\Phi}_{N} - {\Phi}_{N+1} \right) =
 4 (N+1) \int {{\xi}_{N}^2 \over {\xi}_{N+1}^2} \, d \, {\rm log}  {{\xi}_{N}
\over {\xi}_{N+1} } =
\cr
& = 2 (N+1) \left( {{\xi}_{N}
\over {\xi}_{N+1}} \right)^2     \vert_{x_3 = -L}^{x_3 = + L}  =
\quad 2 N {{{\z}_{N-1} {\z}_{N+1}}\over{{\z}_{N}^2}} \vert_{x_3 = -L}^{x_3
= + L}\ ,  \cr
} \  }
and the total charge vanishes as:
\eqn\ttlch{Q = \left[ 2N {{{\z}_{N-1} {\z}_{N+1}}\over{{\z}_{N}^2}}
 +
{\p}_3 \left( {{\z}_{N+1} \over {\z}_{N}} \right)
 \right]_{x_3 = -L}^{x_3 = + L} \equiv 2(N+1)
 \vert_{x_3 = +L} - 2(N+1) \vert_{x_3 = -L} \ . }

{}We can better understand the distribution of the magnetic field by
looking separately at the fluxes through the ``lids'' $x_3 = \pm L$
of our box and through the ``walls'' $n = N$.

{}The walls contribute
$$
\left[ 2N {{{\z}_{N-1} {\z}_{N+1}}\over{{\z}_{N}^2}} \right]_{x_3 = -L}^{x_3
= + L} \sim - {{L}\over{\sqrt{L^2 + N}}} \sim - 1 \ ,
$$
while the lids contribute $\sim + 1$.
Let us isolate the term $B_{3}( +L, n=0) \to +2$ (recall \asm). It contributes
to the flux through the upper lid. The rest of the flux through the lids
is therefore $\sim -1$. Hence the flux
through the rest of the ``sphere at infinity'' is
$-2$ and it is roughly uniformly distributed
($-1$ contribute the walls and $-1$ the lids).  So we get a picture of a
spherical magnetic field of a monopole together with a flux tube
pointing in one direction.

{}This spherical flux becomes observable
in the naive ${\t} \to 0$ limit, in which the string
becomes localized at the point $x_3 = 0$, $n=0$ (since the slope
of the linearly growing ${\Phi}_{0} \sim {x_3 \over {\t}}$ becomes
infinite). In the $\t = 0$ limit
we throw out this point and all of the string.

\newsec{Discussion}

In this paper we found an  explicit
analytic expression for a soliton in the U(1)
gauge theory on a noncommutative space. The solution
describes    a magnetic monopole  attached to a finite
  tension string, that runs off to infinity tranverse to the
noncommutative plane. This soliton has a clear reflection in type
IIB string theory. If the gauge theory is realized as the
${\a}^{\prime} \to 0$ limit of the theory on a D3-brane in the IIB
string theory in the presence of a background NS B-field,
 then the monopole with the string attached is nothing
but the D1-string ending on the D3-brane. What is unusual about the
solution that we found is that it describes this string as a non-singular
field configuration.

Whether this string is a dynamical object in the gauge theory,
with full stringy degrees of freedom,  remains to be determined.
To this end we should analyze the spectrum of the fluctuations of
this string. Several remarks are in order:

\ndt
$\bullet $ First, the ``location'' of the string is not very well defined.,
In noncommutative gauge theory the local energy density, as all local
operators,
is not gauge invariant. However
the energy of a line element of a string,
as a function of $x_3$ is a well-defined
gauge-invariant notion:
\eqn\tnsn{t (x_3) = {1\over{2g_{\rm YM}^2}} \int dx_1 dx_2 \left(
{\vec B}^2 + \left( {\vec \nabla} {\Phi} \right)^2 \right)\ .}
For our solution this ``tension'' turned out to be exponentially small
for $x_3 < 0$ and essentially a constant
\eqn\tnsn{
t = {2{\pi}\over{g_{\rm YM}^2 {\t}}}\ ,}
for $x_3 > 0$.

\ndt $\bullet$ We worked in the gauge where $A_3 = 0$, which still
allows for  $x_3$-independent gauge transformations. This gauge
freedom is broken down to a global $U(1)$ rotation by demanding
that for $x_3 \to + \infty$ $${\Phi}_{0}(x_3) \to -
{{x_3}\over{\t}}\ .$$

\ndt If we impose this asymptotic behaviour on $\Phi$, then the
problem of finding the spectrum of the fluctuations of the string
becomes well-posed.

\ndt $\bullet$ Our solution breaks translational invariance. One
would expect the derivatives ${\p}_{\m} ({\Phi}, A_{c},
A_{c^{\dagger}})$ to show up as zero modes. However, the
derivatives in the $x_1, x_2$ directions are infinitesimal gauge
transformations, while the derivatives in the $x_3$ direction are
not normalizable: $$ {\p}_{c}{\Phi} = [ {\Phi}, c^{\dagger}],
\qquad {\p}_{c} A_{\m} = D_{\m} c^{\dagger} - {\d}_{\m,
c^{\dagger}}$$ (the shift of $A_{\m}$ by a  constant is a symmetry
of the theory).

The next subject which we plan to elaborate further on is the extension of our
analytic solution to the case of $U(2)$ noncommutative gauge theory. In this
case one expects to find   strings of  finite extent, according
to the brane picture \hashimoto.

What is the   relation  between the string we have found and the
electric flux strings found recently in \hklm. These authors also
study the coupled gauge field - Higgs field system,  with the
Higgs field in the adjoint representation. Their Higgs field $t$
arises from the open string tachyon, and has a non-trivial
potential $V(t)$. In the limit of large noncommutativity ${\t}$
the kinetic term can be neglected, according to \gms, and the
soliton can be found as a Gaussian wave-packet localized at the
origin of the transverse plane to the would-be-string space, with
the values of the tachyon field at the origin and far away given
by  at two different critical points of the potential $V(t)$. In
our case we have no potential  for $\Phi$, nor did we assume
${\t}$ to be large. However, our solitonic string also had an
effective thickness of the order of ${\t}$, and also disappears
when ${\t} = 0$. It would be interesting to see, whether S-duality
will  map our magnetic strings to the electric strings of \hklm.

As a step in this direction we would like to compare the tension
of our string with that of D-string (the authors of \hklm\ claim
to have a complete agreement of the tension of their soliton with
the tension of the fundamental string). As already mentioned, a
D-string ending on a D3-brane in the presence of the constant
$B$-field bends. To analyze this bending one could use the exact
solution of the DBI theory \moriyama,  the B-deformed spike
solutions of \curtjuan. However, for our qualitative analysis, it
is sufficient to look at the linearized equations. If we replace
the DBI Lagrangian by its Maxwell approximation, then the BPS
equations in the presence of the $B$-field will have the form:
\eqn\bpsbf{B_{ij} + F_{ij} + \sqrt{{\rm det}g} \, {\ve}_{ijk} \,
g^{kl}{\p}_{l}{\Phi} = 0 \ , } where we should use the closed
string metric \clstr. The solution of \bpsbf\ is:
\eqn\slntn{{\Phi} = B \left( 1 + \left(
{{\t}\over{2{\pi}{\a}^{\prime}}} \right)^2 \right) x_3 -
{1\over{2r}}, \qquad r^2 = x_3^2 + {1\over{\left( 1 + \left(
{{\t}\over{2{\pi}{\a}^{\prime}}} \right)^2 \right)}} \left( x_1^2
+ x_2^2 \right) \ .} The linearly growing piece in ${\Phi}$ should
be interpreted as a global rotation of the D3-brane, by an angle
${\psi}$, ${\rm tan}{\psi} = {{\t}\over{(2{\pi}{\a}^{\prime})}}$.
This conclusion remains correct even after the full non-linear BPS
equation is solved (see \moriyama.  Notice however that we fix
$G_{ij} = {\d}_{ij}$ instead of $g_{ij} = {\d}_{ij}$ as in
\moriyama). The singular part of ${\Phi}$, the spike, represents
the D-string. If we rotate the brane, then the spike forms an
angle ${{\pi}\over{2}} -{\psi}$ with the brane.  If we project
this spike on the brane, then the energy, carried by its shadow
per unit length, is related to the tension of the D-string via:
\eqn\eft{{{T_{D1}}\over{{\rm sin}{\psi}}} =
{1\over{2{\pi}{\a}^{\prime} g_{s}}} {{\sqrt{ (
2{\pi}{\a}^{\prime})^2 + {\t}^2}}\over{{\t}}} = {{(
2{\pi}{\a}^{\prime})^2 + {\t}^2}\over{2{\pi} g_{\rm YM}^2
({\a}^{\prime})^2 {\t}}} \, .} However, this is not the full
story. The endpoint of the D-string is a magnetic charge, which
experiences a constant force, induced by the background magnetic
field. If we had introduced a box of the extent $2L$ in the
$x_3$-direction, then in order to bring a tilted D-string into our
system from outside of the box we would have had to spend an
energy equal to ${T_{D1} \over{{\rm sin}{\psi}}} L$, but we would
have been helped by the magnetic force, which would decrease the
work done by $$ {{2\pi}\over{g_{\rm YM}^2}} B_3 =
{{2\pi}\over{g_{\rm YM}^2}} B_{12} \,\, g^{11} g^{22} \sqrt{g}
\sim {1\over{({\a}^{\prime} g_{\rm YM})^2}} {\t} \, . $$ In sum,
the energy of the semi-infinite D-string in the box will be given
by\foot{We thank K.~Hashimoto for bringing a very helpful argument
from the second reference in \hashimoto\ to our attention}:
\eqn\eft{{{2{\pi}}\over{g_{\rm YM}^2 {\t}}} \, .} This expression
coincides with our tension \tnsn. On dimensional grounds,
non-commutative gauge theory cannot produce any other dependence
of the tension on ${\t}$ but that given  in \tnsn.

Finally, the large ${\t}$ limit of the noncommutative gauge theory
may provide an exciting opportunity to learn more about the large
$N$ non-abelian commutative  gauge theories, for both theories
become essentially planar in this limit. If we keep $g_{\rm YM}^2$
small and take ${\t} \to \infty$ then our magnetic strings become
tensionless. Whether this could  lead to condensation of the
magnetic charges and a mechanism for confinement remains to be
seen.

\listrefs

\bye